\documentclass[prl,nofootinbib,twocolumn,superscriptaddress,preprintnumbers]{revtex4-2}

\usepackage{graphicx}
\usepackage{color}
\usepackage{hepunits}
\usepackage{hyperref}
\usepackage{braket}

\usepackage{graphicx,xfrac}  % needed for figures
\usepackage{dcolumn}   % needed for some tables
\usepackage{bm}        % for math
\usepackage{amssymb, amsmath}
\usepackage{slashed}   % for math

\usepackage{calrsfs}
\DeclareMathAlphabet{\pazocal}{OMS}{zplm}{m}{n}

\usepackage[english]{babel}

\newcommand{\edit}[1]{}

\newcommand{\be}{\begin{eqnarray}}
\newcommand{\ee}{\end{eqnarray}}
\newcommand{\sigmabar}{\overline{\sigma}}

\newcommand{\xhat}{\hat{\mathbf{x}}}
\newcommand{\xx}{\mathbf{x}}
\newcommand{\qq}{\mathbf{q}}
\newcommand{\pp}{\mathbf{p}}
\newcommand{\rr}{\mathbf{r}}
\newcommand{\ahat}{\hat{a}}
\newcommand{\RR}{\mathbf{R}}
\newcommand{\vv}{\mathbf{v}}
\newcommand{\vmin}{v_{\rm min}}
\newcommand{\vminN}{v_{\rm min}^{(n)}}
\newcommand{\omegaO}{\omega_{0}}

\newcommand{\Eq}[1]{Eq.~(\ref{#1})}

\begin{document}

%\vspace{-1cm}
{\small \hfill FERMILAB-PUB-20-588-T}

\title{ Dark Matter Detection With Bound Nuclear Targets: The Poisson Phonon Tail}
%\title{Sub-GeV Dark Matter-Nuclear Scattering in Solid-State Systems: The Poisson Tail}

\author{Yonatan Kahn}
\email{yfkahn@illinois.edu}
\affiliation{Department of Physics, University of Illinois at Urbana-Champaign, Urbana, IL 61801, U.S.A.}
\affiliation{Illinois Center for Advanced Studies of the Universe, University of Illinois at Urbana-Champaign, Urbana, Illinois 61801, U.S.A}

\author{Gordan Krnjaic}
\email{krnjaicg@fnal.gov}
\affiliation{Fermi National Accelerator Laboratory, Batavia, Illinois 60510, U.S.A.}
\affiliation{Kavli Institute for Cosmological Physics, University of Chicago, Chicago, Illinois 60637, U.S.A.}

\author{Bashi Mandava}
\email{mandava3@illinois.edu}
\affiliation{Department of Physics, University of Illinois  at Urbana-Champaign, Urbana, IL 61801, U.S.A.}

\date{\today}

\begin{abstract}
Dark matter (DM) scattering with nuclei in solid-state systems may produce elastic nuclear recoil at high energies and single-phonon excitation at low energies. When the dark matter momentum is comparable to the momentum spread of nuclei bound in a lattice, $q_0 = \sqrt{2 m_N \omegaO}$ where $m_N$ is the mass of the nucleus and $\omegaO$ is the optical phonon energy, an intermediate scattering regime characterized by multi-phonon excitations emerges. We study a greatly simplified model of a single nucleus in a harmonic potential and show that, while the mean energy deposited for a given momentum transfer $q$ is equal to the elastic value $q^2/(2m_N)$, the phonon occupation number follows a Poisson distribution and thus the energy spread is $\Delta E = q\sqrt{\omegaO/(2m_N)}$. This observation suggests that low-threshold calorimetric detectors may have significantly increased sensitivity to sub-GeV DM compared to the expectation from elastic scattering, even when the energy threshold is above the single-phonon energy, by exploiting the tail of the Poisson distribution for phonons above the elastic energy. We use a simple model of electronic excitations to argue that this multi-phonon signal will also accompany ionization signals induced from DM-electron scattering or the Migdal effect. In well-motivated models where DM couples to a heavy, kinetically-mixed dark photon, we show that these signals can probe experimental milestones for cosmological DM production via thermal freeze-out, including the thermal target for Majorana fermion DM.
\end{abstract}

%\vspace{0.3cm}
\maketitle

The strategy of searching for dark matter (DM) by detecting the energy imparted to a nucleus during a scattering process dates back to the original proposals of Goodman and Witten \cite{Goodman:1984dc} and Drukier, Freese, and Spergel \cite{Drukier:1986tm} and forms the basis for the multi-ton liquid noble experiments XENONnT \cite{Aprile:2020vtw} and LZ \cite{Akerib:2019fml}. In recent years, a vast landscape of plausible DM models has opened up at much lighter DM masses compared to the GeV-scale WIMPs these experiments were originally designed to search for (see \cite{Battaglieri:2017aum} and references therein), necessitating a rethinking of the basic assumptions that went into the calculations of the nuclear scattering rates. In particular, for \emph{elastic} scattering of free nuclei, the energy transfer between the DM and nucleus is inefficient if the DM mass is much below the nucleus mass: for a xenon nucleus of mass $m_N = 122 \ {\rm GeV}$ and DM of mass $m_\chi = 100 \ {\rm MeV}$, the typical momentum transfer is $q \sim m_\chi v \sim 150 \ {\rm keV}$, and the recoiling nucleus has energy $q^2/(2m_N) \sim 0.1 \ {\rm eV}$, well below the thresholds of typical noble liquid detectors.

However, as emphasized in recent work \cite{Trickle:2019nya,Griffin:2019mvc}, a nucleus in a lattice is not a free particle, and therefore the elastic scattering argument does not necessarily apply for solid-state detectors. Indeed, an individual nucleus in a solid is subject to a harmonic oscillator potential from its neighboring atoms
\be
\label{eq:VSHO}
V(r) \approx \frac{1}{2} m_N \omegaO^2 r^2,
\ee
where $\omegaO \sim 60 \ {\rm meV}$ is a typical optical phonon energy for silicon (for other materials, $\omegaO \in [ 20 \ {\rm meV}, 140 \ {\rm meV}]$). This potential is short-range, acting only over interatomic distances $a \sim ({\rm keV})^{-1}$, and thus the displacement energy required to remove a nucleus from its lattice site is $E_d \sim \frac{1}{2} m_N \omegaO^2 a^2 \sim \pazocal{O}(10 \ {\rm eV})$. Only for recoil energies $E_R$ well above $E_d$ may the final-state nucleus be well-approximated as a free particle; below $E_d$, the final-state energy spectrum is that of a harmonic oscillator, and of course the initial state is the ground state of a harmonic oscillator rather than a zero-momentum plane wave. This observation has previously been made in the context of single-phonon excitations in solids \cite{Knapen:2017ekk,Griffin:2018bjn}, where DM with de Broglie wavelength exceeding the interatomic spacing interacts primarily with collective modes of many oscillating nuclei, which are quantized into a phonon spectrum.

In this Letter, we study the intermediate regime where the momentum transfer satisfies $q \gg 1/a$, such that the interaction is localized to a single lattice site, but $E_R \lesssim E_d$ so that \emph{both} the initial and final nuclear states belong to the harmonic oscillator spectrum. We construct a toy model of a solid-state detector by considering the non-relativistic single-particle quantum mechanics of a nucleus subject to the potential~(\ref{eq:VSHO}). The characteristic momentum spread of the ground state is
\be
q_0 = \sqrt{2 m_N \omegaO},
\ee
where $q_0 \approx 56 \ {\rm keV}$ for silicon. We will show that when a momentum transfer $q > q_0$ is kinematically allowed, the typical energy deposited is
\be
\overline{E}_R = \left( \frac{q}{q_0} \right)^2 \omegaO = \frac{q^2}{2m_N}.
\ee
This is the energy expected from elastic scattering, but here it may be interpreted as the production of a multi-phonon state with occupation number $n = (q/q_0)^2$. 

We will show that in this greatly simplified model of a single oscillator, the distribution of phonon number is \emph{exactly} Poissonian, so that unlike elastic scattering where only a single energy $E$ is allowed for a given $q$, there is a spread $\Delta n = q/q_0$ corresponding to an energy spread
\be
\Delta E_R = \left( \frac{q}{q_0} \right) \omegaO = q \sqrt{\frac{\omegaO}{2m_N}}.
\ee
This observation extends the reach of solid-state nuclear recoil detectors to sub-GeV DM compared to previous analyses which assumed the kinematics of elastic scattering. As a cross-check, we show that our results parametrically reduce to the previously-calculated single-phonon rate when $q \ll q_0$, offering a pleasing interpretation of the single-phonon rate as an upward Poisson fluctuation when the elastic energy is well below the phonon energy; our results converge exactly on the expected elastic recoil spectrum when $q \gg q_0$. Our work complements that of \cite{Trickle:2019nya,Griffin:2019mvc} by explicitly showing the transition between the single-phonon and elastic regimes in a simple model (see also \cite{Campbell-Deem:2019hdx} which considers production of multiple acoustic phonons for $q \ll q_0$). We will show that when the DM-nucleus interaction is mediated by a heavy dark photon, the full single- and multi-phonon spectrum can probe a wide range of thermal relic targets, including Majorana DM which suffers a velocity-suppressed cross section.

\begin{figure}
\hspace{-0.2in}
\includegraphics[width=2.5in,angle=0]{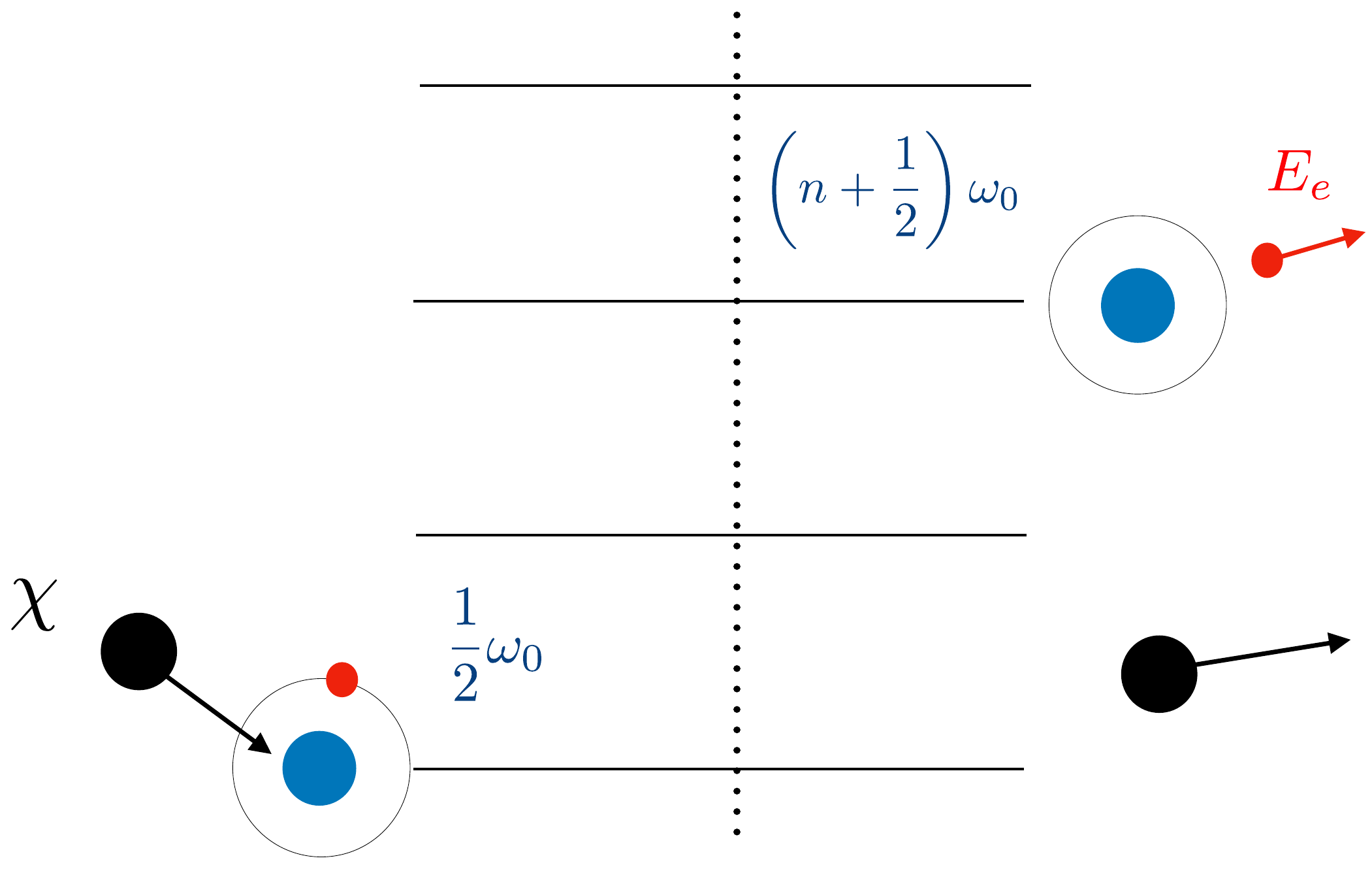}
\caption{ 
Schematic cartoon of DM $\chi$ upscatterng a bound nucleus in a harmonic 
oscillator ground state (left) into an excited state corresponding to the production of $n$ phonons
and possible additional ionization energy $E_e$ from the Migdal effect (right). 
%\vspace{-0.5cm}
 }
\label{Cartoon}
\end{figure}

Furthermore, we argue that when a nuclear recoil induces secondary ionization as in the Migdal effect \cite{Migdal1939,Vergados:2004bm,Moustakidis:2005gx,Bernabei:2007jz,Ibe:2017yqa,Dolan:2017xbu,Bell:2019egg,Baxter:2019pnz,Essig:2019xkx} (see Fig.~\ref{Cartoon} for an illustration), the electron spectrum is independent of the nuclear recoil spectrum except at the very largest kinematically-allowed ionization energies. This result was previously known for isolated atoms \cite{Ibe:2017yqa}, but we show that it persists in a simple model of a bound nucleus.
For similar reasons, we also find that DM-electron scattering also yields secondary phonon excitations. 
 Therefore, the multi-phonon spectrum we compute should be an irreducible component of an ionization signal in solid state detectors, and due to the Poisson tail it may lie above the threshold of next-generation calorimetric detectors and would help to distinguish it from other non-particle backgrounds that create charge pairs without accompanying recoil. Finally, our results suggest a qualitatively different figure of merit for detector materials to detect sub-GeV DM-nuclear scattering: to maximize the sensitivity at low DM masses, large phonon energies are preferred to maximize the Poisson fluctuations when the elastic energy deposit is close to threshold.

\begin{center}
{\bf  Harmonic Oscillator Model}
\end{center}
We consider a single nucleus of mass $m_N$ subject to a 3-dimensional isotropic harmonic oscillator potential:
\be
\hat{H} = \frac{\hat{\pp}_N^2}{2m_N} + \frac{m_N \omegaO^2 }{2} \hat{\rr}_N^2.
\ee
The energy eigenstates are $|\vec{n} \rangle$, where $\vec{n} = \{n_x, n_y, n_z \}$, with energies $E_n = \left(n + \frac{1}{2}\right) \omegaO$, where $n = n_x + n_y + n_z$. To model galactic DM particles $\chi$ scattering from this nucleus, we follow \cite{Essig:2015cda} and write the total scattering rate for a detector with $N_T$ nuclear targets as
\be
\label{eq:Rtot}
R = N_T \frac{\rho_\chi}{m_\chi} \frac{Z^2 \sigmabar_p}{8\pi \mu_{\chi p}^2} \int \frac{d^3 \mathbf  \qq}{q} \sum_n |f(n, \qq)|^2 \eta(\vminN),
\ee
where $\rho_\chi/m_\chi = 0.3 \,{\rm cm}^{-3} ( {\rm GeV}/m_\chi)$ is the local DM number density \cite{deSalas:2019rdi}, $Z$ is the atomic number of the target and $\sigmabar_p$ is a fiducial DM-proton cross section; for concreteness we restrict our attention to a contact interaction between DM and protons, but our results are easily generalizable to other interactions. We do not include a nuclear form factor because the momentum transfers we consider for sub-GeV DM will always be smaller than the inverse nuclear size, $q \ll 1/R_0 \sim \pazocal{O}(10 \ {\rm MeV})$.
% However, in models where the DM couples to charge (for example, through a dark photon mediator, as we will consider later in this Letter), an atomic form factor should be included to account for charge screening at low momentum transfers in monoatomic crystals such as silicon.

%%%%%%%%%%%%%%%%%%
% 	Spectral Figure
%%%%%%%%%%%%%%%%%%

\begin{figure*}
\hspace{-0.2in}
\includegraphics[width=3.3in,angle=0]{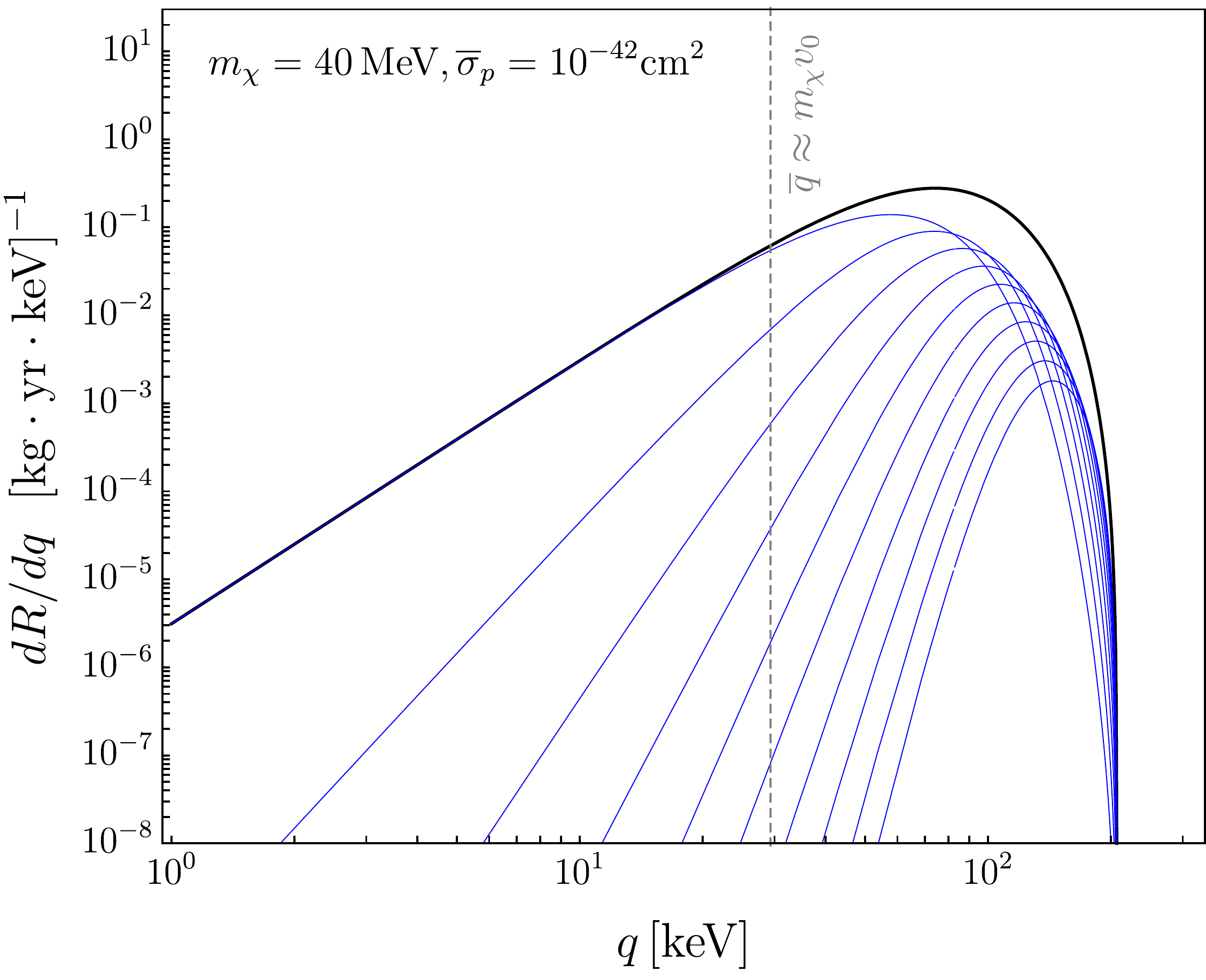}~~
\includegraphics[width=3.3in,angle=0]{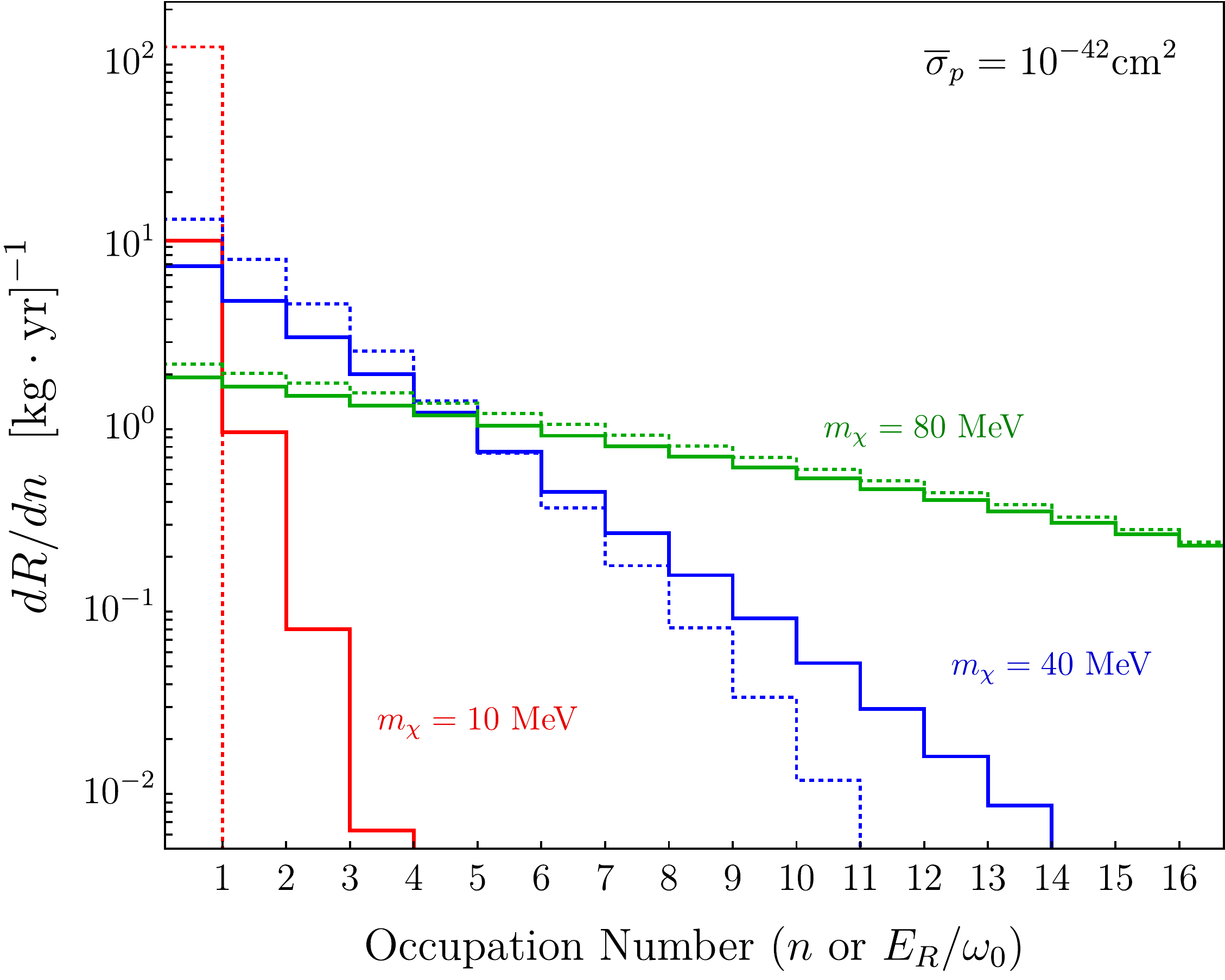}
% \vspace{-0.3cm}
\caption{
({\bf Left:})
Differential momentum spectrum from \Eq{dRdq-avg} with a silicon target. The blue curves represent
contributions from individual $0\to n$ transitions for $n = 1, \dots 10$, with the dotted gray line showing
the typical halo DM momentum $\overline q \approx m_\chi v_0$, where $v_0 = 220$ km/s.
({\bf Right:}) Quantized event rate spectra showing individual $n$ contributions from \Eq{eq:Rtot}
using the angle-averaged Poisson form factor in \Eq{fnq} for various $\chi$ masses. For comparison,
the dotted histograms show the corresponding elastic recoil spectra quantized in units of $E_R/\omega_0$; the effects of the Poisson tail are clearly visible for smaller DM masses.
%\vspace{-0.3cm}
 }
\label{SpectrumPlot}
\end{figure*}

The dimensionless functions appearing in the rate integral are the inelastic form factor
\be
\label{fnq}
|f(n, \qq)|^2 = \sum_{n_x + n_y + n_z = n} | \langle \vec{n} | e^{i \qq \cdot \hat{\rr}_N} | 0 \rangle|^2~,
\ee
and the DM inverse mean speed
\be
\label{eta}
\eta(\vminN) = \int \frac{d^3 \mathbf  \vv}{v} f_\chi(\vv)\,  \Theta \! \left(     v - v_{\rm min}^{(n)}         \right)~,
\ee
where $f_\chi(\vv)$ is the DM velocity distribution, which we assume to be the isotropic Standard Halo Model. Here
\be
\label{eq:vminN}
\vminN = \frac{n \omegaO}{q} + \frac{q}{2m_\chi}~,
\ee
is the minimum DM speed needed to excite the nucleus to harmonic oscillator level $n$.

We can calculate the form factor analytically using either momentum-space wavefunctions or the harmonic oscillator algebra. Due to the isotropy of the harmonic oscillator, the angular integrals can also be performed analytically. Defining an angular average $|f(n,q)|^2 \equiv \frac{1}{4\pi} \int d\Omega_{\qq} |f(n,\qq)|^2$, we find
\be
\label{eq:fNq}
|f(n,q)|^2 = \frac{1}{n!} \left(\frac{q}{q_0}\right)^{2n} \!\! e^{-q^2\!/q_0^2}~.
\ee
While to our knowledge this result has not appeared before in the DM literature, it is known in the case of neutron scattering, see for example \cite{schober2014introduction}.

Performing the angular average in \Eq{eq:Rtot} and using in \Eq{eq:fNq} yields a differential spectrum
\be
\label{dRdq-avg}
\frac{dR}{dq } =  
\! \frac{N_T \rho_\chi}{m_\chi} \frac{Z^2 \sigmabar_p }{2 \mu_{\chi p}^2}
 \sum_n \frac{q}{n!} \left(\frac{q}{q_0}\right)^{2n} \!\!e^{-q^2\!/q_0^2} \,
\eta(v_{\rm min}^{(n)}),~~~
\ee
and in Fig. \ref{SpectrumPlot} we show some representative spectra for different
DM masses. 
See the Supplementary Material for derivations of the main results in this section.

%\begin{center}
%{\bf Poissonian Phonons}
%\end{center}
The form factor in~(\ref{eq:fNq}) is nothing but a Poisson distribution in $n$ with mean $\bar{n} = q^2\!/q_0^2 = q^2/(2 m_N \omegaO)$. Therefore, the mean energy deposited in a DM-nuclear scattering is $\bar{n} \omegaO = q^2/(2m_N)$, the elastic value. In the limit $q^2 \gg q_0^2$, the Poisson distribution approaches a delta function $\delta( n - \bar{n})$; taking the continuum limit  $n \to E_R/\omegaO$, $\sum_n \to \int dn$ in \Eq{eq:Rtot}, 
and replacing $q dq = m_N dE_R$, we recover the usual elastic recoil spectrum 
\be
\label{dRdER_elastic}
\frac{dR}{dE_R} \approx N_T m_N \frac{\rho_\chi}{m_\chi}\frac{Z^2 \sigmabar_p}{2 \mu_{\chi p}^2} \, \eta(\vmin^{(\bar n)}),
\ee
with $\vmin^{(\bar n)} = \sqrt{m_N E_R/(2\mu_{\chi N}^2)}$. This matching justifies the standard approximation of treating the nucleus as a free particle with the elastic dispersion relation, at least for $\omegaO \ll E_R < E_d$ such that the nucleus remains bound in the harmonic oscillator potential.

Consider now the opposite limit, $q^2 \ll q_0^2$. From \cite{Knapen:2017ekk,Griffin:2018bjn,Trickle:2019nya,Griffin:2019mvc}, the form factor for production of a single optical phonon is parametrically $(q/q_0)^2 e^{-q^2\! /q_0^2}$, where in the context of condensed matter physics the exponential is known as the (zero-temperature) Debye-Waller factor. We can now interpret this as the Poisson probability for $n = 1$ phonons when $\bar{n} \ll 1$. While the highest-probability outcome is producing no phonons, $n = 0$, the most likely excitation above the ground state is $n = 1$, with larger phonon numbers strongly suppressed by powers of $q^2 \! /q_0^2 \ll 1$.

The advantage of our model is that we can now seamlessly interpolate between the single-phonon regime $q^2 \ll q_0^2$ and the elastic regime $q^2 \gg q_0^2$. When $q^2 \sim q_0^2$, as is the case for kinematics of 50 MeV DM in the Standard Halo Model, the Poissonian fluctuations in phonon number become important. In the right panel of Fig. \ref{SpectrumPlot} we show representative quantized spectra that illustrate this behavior. Indeed, consider a detector with threshold close to $\omegaO$. If the scattering were purely elastic, DM with maximum momentum $q_0$ could not produce a detectable nuclear recoil. However, for $q = q_0$, the Poisson probability for $n = 2$ phonons is 0.18, compared to 0.37 for $n = 1$. Thus the true reach of the detector extends to lower DM masses, because the probability to deposit energy well above threshold is comparable to the probability to deposit energy at threshold. A similar version of this argument applies to detectors with thresholds somewhat above the optical phonon energy: the Poisson tail of events with large $n$ permits sensitivity to smaller DM masses than would be expected based on elastic kinematics alone.

\begin{center}
{\bf  Nuclear Scattering Reach} 
\end{center}

%%%%%%%%%%%%%%%%%%
% 		Reach Figure
%%%%%%%%%%%%%%%%%%

\begin{figure}
\hspace{-0.2in}
\includegraphics[width=3.5in,angle=0]{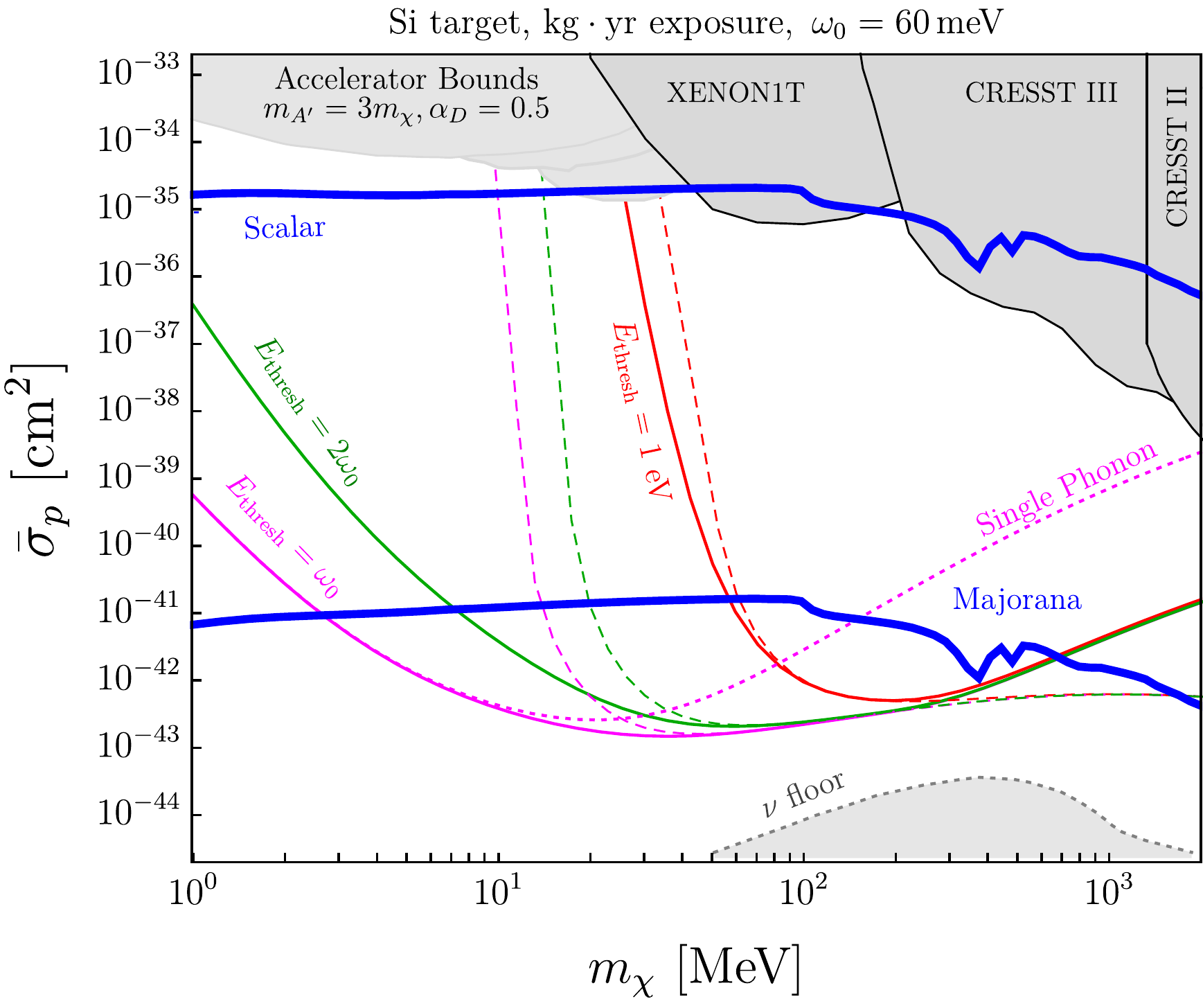}
%\vspace{-0.4cm}
\caption{ 
Exclusion curves (3 events, zero background) for various
 phonon detection thresholds computed using \Eq{eq:Rtot} with 
 $n \leq E_{\rm thresh}/\omegaO$ (solid), along with the equivalent curves for elastic scattering (long dashed) and single-phonon production (short dashed magenta). 
 The gray shaded regions represent the accelerator bounds on invisibly decaying 
$A^\prime$  from the LSND \cite{deNiverville_2011}, NA64 \cite{NA64:2019imj}, E137 \cite{Batell:2014mga,PhysRevLett.121.041802}, BABAR \cite{Izaguirre:2013uxa,Essig:2013vha,Lees:2017lec}, and MiniBooNE \cite{Aguilar_Arevalo_2020} experiments. Also shown are limits from
XENON1T \cite{Aprile:2019xxb} (limits on electron scattering interpreted as $\bar{\sigma}_p$ in the dark photon model),
CRESST II \cite{PhysRevD.100.102002}, and
 CRESST III \cite{Abdelhameed:2019hmk}. We also show the 
 neutrino floor background from  \cite{Battaglieri:2017aum}.
%\vspace{-0.3cm}
 }
\label{ReachPlot}
\end{figure}

Now we apply our observations to estimate the 
reach of a next-generation calorimetric detector whose
sensitivity to light DM can be greatly enhanced 
by including {\it irreducible} multi-phonon contributions from the
Poisson tail. For concreteness, we consider a DM-nucleus interaction mediated by a 
massive, kinetically-mixed dark photon $A^\prime$ with Lagrangian
\be
{\cal L}_{\rm int} = A^\prime_\mu \left(  \epsilon e J_{\rm EM}^\mu + g_D J_{D}^\mu \right),
\ee
where $\epsilon$ is a kinetic mixing parameter, $g_D$ is the
$A^\prime$-$\chi$ coupling constant,
 $J^\mu_{\rm EM}$ is the electromagnetic current,
  and $J^\mu_D$ is the dark matter current.  
In the contact limit $m_{A^\prime} \gg q$, for complex scalar $\chi$, the single proton
cross section is  
 \be
 \label{sigmabar}
\overline \sigma_p = 
\frac{16 \pi \epsilon^2 \alpha \alpha_D \mu_{\chi p}^2}{m_{A^\prime}^4}~,
 \ee
 which enters into \Eq{eq:Rtot}. For $m_{A^\prime} > m_\chi$, this
  model can realize thermal freeze-out via $\chi \chi^* \to f^+f^-$ annihilation where $f$ is a Standard Model fermion,
  and there is a one-to-one correspondence between
  $\overline \sigma_p$ and the early universe annihilation rate \cite{Izaguirre:2015yja,Berlin_2019}.
  
In Fig. \ref{ReachPlot} we show how including the additional phonons from
the Poisson tail can enhance the DM signal yield for detectors with various energy thresholds, from 1 eV down to the single-phonon energy $\omegaO$. Each solid curve represents
a different threshold for a silicon crystal target. We only include transitions
into bound oscillator final states $0 \to n$ with $n \omegaO < E_d$, so this simple model 
represents a lower bound on the total signal rate. The dashed curves show the
would-be sensitivity of a low-threshold detector in the elastic
regime, demonstrating that our result continuously interpolates between 
the discrete phonon regime and elastic DM-nuclear scattering which
takes over for $\bar{n} \gg 1$, and that multi-phonon production provides increased sensitivity at lower DM masses. Note that, as in \cite{Griffin:2019mvc}, the dashed elastic curves in this figure flatten towards higher masses reflecting the $\mu_{\chi p}$ dependence in \Eq{dRdER_elastic}; the inelastic phonon curves shift upwards as a greater fraction
of halo particles can deposit energy above $E_d$ to displace the nucleus, which we do not consider in our model. 

Here, the $A^\prime$ couples universally to charge, so we also include an atomic form factor
in \Eq{eq:Rtot} to account for screening
 \cite{Schiff:1951zza,Tsai:1973py,Emken:2019tni}:
\be
|F_A(q)|^2 = \frac{(\lambda_{\rm TF}^2 q^2)^2}{(1 + \lambda_{\rm TF}^2 q^2)^2}~,
\ee
where $\lambda_{\rm TF} \approx 0.89/Z^{1/3} a_0 \approx 0.37 a_0$ is the Thomas-Fermi screening length for silicon and $a_0$ is the Bohr radius. At low momentum $q \ll 1/\lambda_{\rm TF}$, $|F_A(q)|^2  \to 0$, reflecting complete charge screening by the neutral atom. This is also a manifestation of the well-known fact that dark photons do not couple efficiently to optical phonons at low momentum in non-polar crystals, since out-of-phase oscillations are suppressed compared to in-phase oscillations by powers of $q$. However, because $\lambda_{\rm TF} q_0\approx 5.2$ in silicon, $|F_A(q_0)|^2 = 0.93$, so this screening only affects the kinematic regime with $q \lesssim q_0$. In practice, screening slightly suppresses the single-phonon rate and the total rate below $m_\chi \simeq 10 \ {\rm MeV}$ compared to a generic heavy mediator coupling only to protons. For larger DM masses, our single-phonon curve roughly reproduces the analysis of \cite{Griffin:2019mvc} for a heavy hadrophilic mediator, which is an important check on the validity of this simple model. 

The blue curves in Fig. \ref{ReachPlot} correspond to thermal freeze-out targets for complex scalar and Majorana fermion DM candidates coupled to
$A^\prime$. The Majorana cross section is proportional to $\overline \sigma_p$ in \Eq{sigmabar} (which
is defined for scalar $\chi$), and further suppressed by $v^2 \sim 10^{-6}$,
where $v$ is the DM velocity in the earth frame (see \cite{Berlin_2019} for details). Both models here feature $p$-wave DM annihilation
and are, therefore, safe from CMB bounds, which exclude
freeze-out for $s$-wave candidates with $m_\chi \lesssim 10$ GeV \cite{Aghanim:2018eyx}. It is notable that 
calorimeters with sufficiently low thresholds in the few-phonon range
 may soon begin to explore the Majorana thermal target.

\begin{center}
{\bf  Migdal Effect and Electron Scattering} 
\end{center}
Nearly a century ago it was pointed out by Migdal \cite{Migdal1939} that a sudden impulse delivered to the nucleus could result in electronic transitions in atoms. In recent years this observation has become increasingly relevant for light DM searches because the ionization energy may greatly exceed the elastic energy (for calorimetric detectors) or quenched elastic energy (for ionization detectors) from nuclear recoil, improving the possibility of detecting nuclear recoil events through secondary ionization even if the nuclear recoil energy is below the detector threshold.

 However, as noted in \cite{Baxter:2019pnz}, the theoretical formalism of the Migdal effect only applies to isolated atoms, where the energy eigenstates are free-particle plane waves and thus a unitary transformation may be applied to transform to the frame of the recoiling nucleus. The overlap between this frame and the stationary frame leads to a factor of $\exp(i \qq_e \cdot \rr_e)$ in matrix elements between electronic states, where $\qq_e \equiv (m_e/m_N)\qq$ is suppressed from the physical momentum transfer $\qq$ by the ratio of electron to nuclear masses. Despite the rather limited validity of the formalism, theoretical estimates of Migdal rates have been performed for semiconductors \cite{Ibe:2017yqa}, including for the valence bands \cite{Essig:2019xkx}, where the effective mass of the electron may differ considerably from the vacuum mass $m_e$ used in the standard Migdal calculation. 

While we do not attempt to provide a rigorous derivation of the Migdal effect in solid-state systems in this paper (especially because many-body electron states such as the plasmon may contribute significantly \cite{Kurinsky:2020dpb,Kozaczuk:2020uzb}), we consider a simplified model in the spirit of our nuclear scattering analysis, a nucleus in a harmonic potential which exerts a potential $V_e$ on a single electron:
\be
\hat{H} = \frac{\hat{\pp}_N^2}{2 m_N} + \frac{\hat{\pp}_e^2}{2m_e} +\frac{m_N \omegaO^2 }{2} \hat{\rr}_N^2 +V_e(\hat{\rr}_N - \hat{\rr}_e).
\ee
Transforming to relative and center-of-mass coordinates, $\hat{\rr} \equiv \hat{\rr}_N - \hat{\rr}_e$, $\hat{\RR} =  (m_N \hat{\rr}_N+  m_e\hat{\rr}_e)/(m_N + m_e)$, the Hamiltonian becomes separable up to small perturbations (see the Supplementary Material), 
and the eigenstates can be written $|\Psi \rangle = | \vec{n};  \psi_e \rangle$, where the 
first label is the harmonic oscillator state and $\psi_e(\rr)$ is the electronic 
wavefunction. The generalized transition amplitudes from \Eq{fnq} now 
factorize into oscillator and ionization terms
\be 
\langle \Psi^\prime | e^{i \qq \cdot \hat{\rr}_N} |\Psi \rangle \simeq \langle \vec{n}| e^{i \qq \cdot \hat{\RR}} | 0 \rangle \langle \psi_{e}^\prime | e^{i \qq_e \cdot \hat{\rr}} | \psi_e \rangle~~~~~~
\ee
where $\qq_e \equiv (m_e/m_N)\qq$ is the familiar Migdal
ionization factor \cite{Ibe:2017yqa}.
If the ionization matrix element depends only on the magnitude $|\qq_e|$, the differential scattering rate factorizes
\be
\label{dRmigdal}
\frac{dR}{dq dE_e} \! =  
\! \frac{N_T \rho_\chi}{m_\chi} \frac{Z^2 \sigmabar_p }{2 \mu_{\chi p}^2}
\! \sum_n q |f(n, q)|^2 
\!\sum_f  \! \frac{d|Z_f|^2}{dE_e}
\eta(v_{\rm min}^{(n,e)}),~~~~~~
\ee
% the Hamiltonian becomes $\hat{H} = \hat{H}_0 + \Delta \hat{H}$, with 
%\be
%\hat{H}_0  &=& \frac{\hat{\pp}_R^2}{2(m_N+m_e)} + \frac{\hat{\pp_r}^2}{2\mu} +\frac{m_N \omegaO^2}{2}  \hat{\RR}^2 -V_e(\hat{\rr}),~~
%\ee
%and expanding the harmonic term gives
%\be
%\Delta \hat{H}  &=& -\mu \omegaO^2 \hat{\RR} \cdot \hat{\rr} +  \frac{\mu^2 \omegaO^2}{2m_N}\, \hat{\rr}^2~,
%\ee
%where $\mu = m_e m_N(m_e + m_N) \approx m_e$ is the electron-nucleus reduced mass. 
%
%Written in this way, $\hat{H}_0$ is separable and can be solved by $\Psi(\RR, \rr) = \psi_N(\RR) \psi_e(\rr)$, where $\psi_N$ is a simple harmonic oscillator wavefunction for the nucleus $N$ and $\psi_e$ an electronic wavefunction. The terms in $\Delta \hat{H}$ are suppressed by powers of $m_e/m_N \ll 1$ and can be treated as a perturbation. In particular, the first-order energy shift is parametrically
%\be
%\label{eq:Eshift}
%\Delta E = \langle \Psi | \Delta \hat{H} | \Psi \rangle \sim \frac{m_e^2}{m_N}\omegaO^2 a^2~,
%\ee
%where we have assumed that the typical spread in position space of the electronic wavefunction is of order the lattice spacing $a$; note that $\Delta E$ independent of the harmonic oscillator level $N$ because $\langle \hat{\RR} \rangle = 0$ in any stationary state, so only the second term in $\Delta \hat{H}$ contributes. For $\omegaO \sim 50 \ {\rm meV}$ and $m_N = 26 \ {\rm GeV}$ as for silicon, we have $\Delta E \sim 25 \ {\rm neV} \ll \omegaO$ and thus we are justified in ignoring the perturbation and treating the nuclear spectrum as purely a harmonic oscillator spectrum.
where $f(n,q)$ is the Poisson distribution in \Eq{eq:fNq} and
\be
|Z_f|^2 \equiv  |\langle   \psi_e^\prime | e^{i\qq_e \cdot \hat \rr}  | \psi_e \rangle |^2
\ee
is the ionization probability as in \cite{Ibe:2017yqa}; if $Z_f$ does not factorize, the
integration is a convolution of the ionization and oscillator form factors and may deviate from the Poisson form (see Supplementary Material). 

We see that in this model, the electronic spectrum is unmodified to leading order in $m_e/m_N$ compared to the free-nucleus picture, and the only changes come in replacing the initial and final nuclear states with harmonic oscillator states, the matrix elements of which we have already calculated. Since the matrix element factorizes, the \emph{only} coupling between the electronic and nuclear excitation energies in \Eq{dRmigdal} comes from energy conservation, which after integrating over the DM velocity distribution modifies $\vminN$ in~\Eq{eq:vminN} to
\be
v_{\rm min}^{(n,e)} = \frac{E_e + n \omegaO}{q} + \frac{q}{2m_\chi}.
\ee
As our earlier analysis has shown, the mean nuclear energy is the elastic energy, which is typically much less than the electronic excitation energy, and thus the only effect of the harmonic oscillator spectrum in this model is to truncate the electronic spectrum at slightly smaller energies than would be expected from elastic nuclear scattering due to the Poisson tail. Compared to previous results on isolated atoms \cite{Ibe:2017yqa}, our new result is the Poisson spectrum of phonons which replaces the continuum of elastic recoil energies for a free nucleus.

 On the other hand, for DM scattering directly off an electron \cite{Essig:2011nj,Essig:2015cda}, the matrix element is proportional to $\langle \Psi^\prime | e^{i \qq \cdot \hat{\rr}_e} |\Psi \rangle$, which now involes $\rr_e = \RR - \frac{\mu}{m_e}\rr$, where $\mu \equiv m_e m_N/(m_e + m_N) \approx m_e$. In the limit $m_N \gg m_e$, the above analysis yields 
\be 
\langle \Psi^\prime | e^{i \qq \cdot \hat{\rr}_e} |\Psi \rangle \simeq \langle \vec{n} | e^{i \qq \cdot \hat{\RR}} | 0 \rangle \langle \psi_{e}^\prime | e^{-i \qq \cdot \hat{\rr}} | \psi_e \rangle~~~~~
\ee
where the electron matrix element, $\langle \psi'_e | e^{-i \qq \cdot \hat{\rr}} | \psi_e \rangle$, is familiar from previous analyses of  DM-electron scattering, but the nuclear matrix element is identical to that of the the Migdal effect. Note that unlike in the case of the Migdal effect, since the coefficient of $\rr$ in $\rr_e$ is unity up to  $\pazocal{O}(m_e/m_N)$, the electron scattering matrix element is independent of the vacuum electron mass to leading order and can be computed entirely with band-structure wavefunctions which fully take into account the effective mass.

Our arguments about electronic excitations are fairly general but rely on an electronic potential which is a function only of $\rr_N - \rr_e$ and a system which is separable enough that it can be treated as a two-body problem and decomposed into relative and center-of-mass coordinates. At this point it is unclear to us how well these assumptions hold in realistic solid-state systems (at the very least, this analysis completely ignores electron-phonon interactions and anisotropies due to the lattice structure which are likely to couple the two spectra at some level beyond simple kinematics), but regardless, our analysis of the nuclear scattering matrix element suggests that the Poisson tail may help push the nuclear recoil energies above threshold. A detector capable of converting ionization energy to phonons, such as the low-threshold calorimetric detectors used by SuperCDMS \cite{Kurinsky:2016fvj,Agnese:2018col,Alkhatib:2020slm} and EDELWEISS \cite{Armengaud:2019kfj,Arnaud:2020svb}, could observe the Poisson phonon spectrum of nuclear recoil simultaneously with the electron-hole pairs created by ionization, which could help distinguish between true scattering events and low-momentum-transfer background processes such as charge leakage.

\begin{center}
{\bf Conclusions} 
\end{center}
 In this Letter we have constructed a simple quantum-mechanical model which describes the multi-phonon regime of DM-nuclear scattering in solid-state systems, interpolating between the elastic regime and single-phonon production when the energy deposit is less than the displacement energy $E_d$. Our key finding is that when the elastic energy $q^2/2m_N$ is close to the optical phonon energy $\omegaO$, there is an order-1 variance in the number of phonons produced, such that the sensitivity of low-threshold detectors to low-mass DM is stronger than previously expected from elastic kinematics alone. This observation suggests that detectors with an especially large optical phonon energy, such as diamond \cite{Kurinsky:2019pgb}, may be able to take advantage of these Poisson fluctuations even when the detector threshold is somewhat above the single-phonon energy.
 
 As an illustrative example, we have shown that plausible next-generation calorimeter detectors with $\sim$ eV scale
 thresholds can exploit this irreducible effect to greatly enhance their sensitivity
 to sub-GeV DM. Remarkably, the potential gains identified here could enable such detectors to probe the full thermal
relic parameter space for Majorana DM candidates with mass between 10 MeV and 1 GeV freezing out via kinetically-mixed dark photons, corresponding to parameter space which is far below current direct detection limits. 
 
 Finally, according to \cite{schober2014introduction}, a more realistic model would incorporate a nontrivial phonon density of states and anharmonicities which broaden the phonon spectrum, but we expect that the parametric scaling of our result would persist, at least for monoatomic crystals. Crystals with more than one atom per unit cell, in particular polar materials where dark photons couple to optical phonons even at low $q$ \cite{Knapen:2017ekk,Griffin:2018bjn,Trickle:2019nya,Griffin:2019mvc}, may exhibit similar behavior but deserve a dedicated analysis, especially because of their directional detection capabilities. We emphasize here that the rates we have computed are only \emph{lower} bounds on the total nuclear scattering rates for sub-GeV DM. Indeed, DM heavier than about 10 MeV has sufficient kinetic energy to displace a nucleus from its lattice site, and the fact that neither the initial nor the final states are free plane waves may allow for the possibility of inelastic scattering when the elastic rate below threshold is zero. This would give a spectrum of recoil events with $E_N > E_d$ additive to the one we consider here. We plan to investigate this possibility in future work.

\medskip
{\it Acknowledgments. --- We thank Daniel Baxter, Gordon Baym, Simon Knapen, Jonathan Kozaczuk, Noah Kurinsky, and Tongyan Lin for many enlightening discussions. The work of YK is supported in part by US Department of Energy grant DE-SC0015655. 
 This manuscript has been authored by Fermi Research Alliance, LLC under Contract No. DE-AC02-07CH11359 with the U.S. Department of Energy, Office of High Energy Physics.
} 

\medskip
{\it 
Note added --- In the final stages of prepararing this work, \cite{toappear} appeared, which 
complements this Letter with a study of the Migdal effect in realistic semiconductor materials and the associated multi-phonon response.
} 

\bibliography{MigdalPhononBib}

%merlin.mbs apsrev4-1.bst 2010-07-25 4.21a (PWD, AO, DPC) hacked
%Control: key (0)
%Control: author (8) initials jnrlst
%Control: editor formatted (1) identically to author
%Control: production of article title (-1) disabled
%Control: page (0) single
%Control: year (1) truncated
%Control: production of eprint (0) enabled
\begin{thebibliography}{49}%
\makeatletter
\providecommand \@ifxundefined [1]{%
 \@ifx{#1\undefined}
}%
\providecommand \@ifnum [1]{%
 \ifnum #1\expandafter \@firstoftwo
 \else \expandafter \@secondoftwo
 \fi
}%
\providecommand \@ifx [1]{%
 \ifx #1\expandafter \@firstoftwo
 \else \expandafter \@secondoftwo
 \fi
}%
\providecommand \natexlab [1]{#1}%
\providecommand \enquote  [1]{``#1''}%
\providecommand \bibnamefont  [1]{#1}%
\providecommand \bibfnamefont [1]{#1}%
\providecommand \citenamefont [1]{#1}%
\providecommand \href@noop [0]{\@secondoftwo}%
\providecommand \href [0]{\begingroup \@sanitize@url \@href}%
\providecommand \@href[1]{\@@startlink{#1}\@@href}%
\providecommand \@@href[1]{\endgroup#1\@@endlink}%
\providecommand \@sanitize@url [0]{\catcode `\\12\catcode `\$12\catcode
  `\&12\catcode `\#12\catcode `\^12\catcode `\_12\catcode `\%12\relax}%
\providecommand \@@startlink[1]{}%
\providecommand \@@endlink[0]{}%
\providecommand \url  [0]{\begingroup\@sanitize@url \@url }%
\providecommand \@url [1]{\endgroup\@href {#1}{\urlprefix }}%
\providecommand \urlprefix  [0]{URL }%
\providecommand \Eprint [0]{\href }%
\providecommand \doibase [0]{http://dx.doi.org/}%
\providecommand \selectlanguage [0]{\@gobble}%
\providecommand \bibinfo  [0]{\@secondoftwo}%
\providecommand \bibfield  [0]{\@secondoftwo}%
\providecommand \translation [1]{[#1]}%
\providecommand \BibitemOpen [0]{}%
\providecommand \bibitemStop [0]{}%
\providecommand \bibitemNoStop [0]{.\EOS\space}%
\providecommand \EOS [0]{\spacefactor3000\relax}%
\providecommand \BibitemShut  [1]{\csname bibitem#1\endcsname}%
\let\auto@bib@innerbib\@empty
%</preamble>
\bibitem [{\citenamefont {Goodman}\ and\ \citenamefont
  {Witten}(1985)}]{Goodman:1984dc}%
  \BibitemOpen
  \bibfield  {author} {\bibinfo {author} {\bibfnamefont {M.~W.}\ \bibnamefont
  {Goodman}}\ and\ \bibinfo {author} {\bibfnamefont {E.}~\bibnamefont
  {Witten}},\ }\href {\doibase 10.1103/PhysRevD.31.3059} {\bibfield  {journal}
  {\bibinfo  {journal} {Phys. Rev. D}\ }\textbf {\bibinfo {volume} {31}},\
  \bibinfo {pages} {3059} (\bibinfo {year} {1985})}\BibitemShut {NoStop}%
\bibitem [{\citenamefont {Drukier}\ \emph {et~al.}(1986)\citenamefont
  {Drukier}, \citenamefont {Freese},\ and\ \citenamefont
  {Spergel}}]{Drukier:1986tm}%
  \BibitemOpen
  \bibfield  {author} {\bibinfo {author} {\bibfnamefont {A.}~\bibnamefont
  {Drukier}}, \bibinfo {author} {\bibfnamefont {K.}~\bibnamefont {Freese}}, \
  and\ \bibinfo {author} {\bibfnamefont {D.}~\bibnamefont {Spergel}},\ }\href
  {\doibase 10.1103/PhysRevD.33.3495} {\bibfield  {journal} {\bibinfo
  {journal} {Phys. Rev. D}\ }\textbf {\bibinfo {volume} {33}},\ \bibinfo
  {pages} {3495} (\bibinfo {year} {1986})}\BibitemShut {NoStop}%
\bibitem [{\citenamefont {Aprile}\ \emph {et~al.}(2020)\citenamefont {Aprile}
  \emph {et~al.}}]{Aprile:2020vtw}%
  \BibitemOpen
  \bibfield  {author} {\bibinfo {author} {\bibfnamefont {E.}~\bibnamefont
  {Aprile}} \emph {et~al.} (\bibinfo {collaboration} {XENON}),\ }\href@noop {}
  {\  (\bibinfo {year} {2020})},\ \Eprint {http://arxiv.org/abs/2007.08796}
  {arXiv:2007.08796 [physics.ins-det]} \BibitemShut {NoStop}%
\bibitem [{\citenamefont {Akerib}\ \emph {et~al.}(2020)\citenamefont {Akerib}
  \emph {et~al.}}]{Akerib:2019fml}%
  \BibitemOpen
  \bibfield  {author} {\bibinfo {author} {\bibfnamefont {D.}~\bibnamefont
  {Akerib}} \emph {et~al.} (\bibinfo {collaboration} {LZ}),\ }\href {\doibase
  10.1016/j.nima.2019.163047} {\bibfield  {journal} {\bibinfo  {journal} {Nucl.
  Instrum. Meth. A}\ }\textbf {\bibinfo {volume} {953}},\ \bibinfo {pages}
  {163047} (\bibinfo {year} {2020})},\ \Eprint
  {http://arxiv.org/abs/1910.09124} {arXiv:1910.09124 [physics.ins-det]}
  \BibitemShut {NoStop}%
\bibitem [{\citenamefont {Battaglieri}\ \emph {et~al.}(2017)\citenamefont
  {Battaglieri} \emph {et~al.}}]{Battaglieri:2017aum}%
  \BibitemOpen
  \bibfield  {author} {\bibinfo {author} {\bibfnamefont {M.}~\bibnamefont
  {Battaglieri}} \emph {et~al.},\ }in\ \href@noop {} {\emph {\bibinfo
  {booktitle} {{U.S. Cosmic Visions: New Ideas in Dark Matter}}}}\ (\bibinfo
  {year} {2017})\ \Eprint {http://arxiv.org/abs/1707.04591} {arXiv:1707.04591
  [hep-ph]} \BibitemShut {NoStop}%
\bibitem [{\citenamefont {Trickle}\ \emph {et~al.}(2020)\citenamefont
  {Trickle}, \citenamefont {Zhang}, \citenamefont {Zurek}, \citenamefont
  {Inzani},\ and\ \citenamefont {Griffin}}]{Trickle:2019nya}%
  \BibitemOpen
  \bibfield  {author} {\bibinfo {author} {\bibfnamefont {T.}~\bibnamefont
  {Trickle}}, \bibinfo {author} {\bibfnamefont {Z.}~\bibnamefont {Zhang}},
  \bibinfo {author} {\bibfnamefont {K.~M.}\ \bibnamefont {Zurek}}, \bibinfo
  {author} {\bibfnamefont {K.}~\bibnamefont {Inzani}}, \ and\ \bibinfo {author}
  {\bibfnamefont {S.}~\bibnamefont {Griffin}},\ }\href {\doibase
  10.1007/JHEP03(2020)036} {\bibfield  {journal} {\bibinfo  {journal} {JHEP}\
  }\textbf {\bibinfo {volume} {03}},\ \bibinfo {pages} {036} (\bibinfo {year}
  {2020})},\ \Eprint {http://arxiv.org/abs/1910.08092} {arXiv:1910.08092
  [hep-ph]} \BibitemShut {NoStop}%
\bibitem [{\citenamefont {Griffin}\ \emph {et~al.}(2020)\citenamefont
  {Griffin}, \citenamefont {Inzani}, \citenamefont {Trickle}, \citenamefont
  {Zhang},\ and\ \citenamefont {Zurek}}]{Griffin:2019mvc}%
  \BibitemOpen
  \bibfield  {author} {\bibinfo {author} {\bibfnamefont {S.~M.}\ \bibnamefont
  {Griffin}}, \bibinfo {author} {\bibfnamefont {K.}~\bibnamefont {Inzani}},
  \bibinfo {author} {\bibfnamefont {T.}~\bibnamefont {Trickle}}, \bibinfo
  {author} {\bibfnamefont {Z.}~\bibnamefont {Zhang}}, \ and\ \bibinfo {author}
  {\bibfnamefont {K.~M.}\ \bibnamefont {Zurek}},\ }\href {\doibase
  10.1103/PhysRevD.101.055004} {\bibfield  {journal} {\bibinfo  {journal}
  {Phys. Rev. D}\ }\textbf {\bibinfo {volume} {101}},\ \bibinfo {pages}
  {055004} (\bibinfo {year} {2020})},\ \Eprint
  {http://arxiv.org/abs/1910.10716} {arXiv:1910.10716 [hep-ph]} \BibitemShut
  {NoStop}%
\bibitem [{\citenamefont {Knapen}\ \emph {et~al.}(2018)\citenamefont {Knapen},
  \citenamefont {Lin}, \citenamefont {Pyle},\ and\ \citenamefont
  {Zurek}}]{Knapen:2017ekk}%
  \BibitemOpen
  \bibfield  {author} {\bibinfo {author} {\bibfnamefont {S.}~\bibnamefont
  {Knapen}}, \bibinfo {author} {\bibfnamefont {T.}~\bibnamefont {Lin}},
  \bibinfo {author} {\bibfnamefont {M.}~\bibnamefont {Pyle}}, \ and\ \bibinfo
  {author} {\bibfnamefont {K.~M.}\ \bibnamefont {Zurek}},\ }\href {\doibase
  10.1016/j.physletb.2018.08.064} {\bibfield  {journal} {\bibinfo  {journal}
  {Phys. Lett. B}\ }\textbf {\bibinfo {volume} {785}},\ \bibinfo {pages} {386}
  (\bibinfo {year} {2018})},\ \Eprint {http://arxiv.org/abs/1712.06598}
  {arXiv:1712.06598 [hep-ph]} \BibitemShut {NoStop}%
\bibitem [{\citenamefont {Griffin}\ \emph {et~al.}(2018)\citenamefont
  {Griffin}, \citenamefont {Knapen}, \citenamefont {Lin},\ and\ \citenamefont
  {Zurek}}]{Griffin:2018bjn}%
  \BibitemOpen
  \bibfield  {author} {\bibinfo {author} {\bibfnamefont {S.}~\bibnamefont
  {Griffin}}, \bibinfo {author} {\bibfnamefont {S.}~\bibnamefont {Knapen}},
  \bibinfo {author} {\bibfnamefont {T.}~\bibnamefont {Lin}}, \ and\ \bibinfo
  {author} {\bibfnamefont {K.~M.}\ \bibnamefont {Zurek}},\ }\href {\doibase
  10.1103/PhysRevD.98.115034} {\bibfield  {journal} {\bibinfo  {journal} {Phys.
  Rev. D}\ }\textbf {\bibinfo {volume} {98}},\ \bibinfo {pages} {115034}
  (\bibinfo {year} {2018})},\ \Eprint {http://arxiv.org/abs/1807.10291}
  {arXiv:1807.10291 [hep-ph]} \BibitemShut {NoStop}%
\bibitem [{\citenamefont {Campbell-Deem}\ \emph {et~al.}(2020)\citenamefont
  {Campbell-Deem}, \citenamefont {Cox}, \citenamefont {Knapen}, \citenamefont
  {Lin},\ and\ \citenamefont {Melia}}]{Campbell-Deem:2019hdx}%
  \BibitemOpen
  \bibfield  {author} {\bibinfo {author} {\bibfnamefont {B.}~\bibnamefont
  {Campbell-Deem}}, \bibinfo {author} {\bibfnamefont {P.}~\bibnamefont {Cox}},
  \bibinfo {author} {\bibfnamefont {S.}~\bibnamefont {Knapen}}, \bibinfo
  {author} {\bibfnamefont {T.}~\bibnamefont {Lin}}, \ and\ \bibinfo {author}
  {\bibfnamefont {T.}~\bibnamefont {Melia}},\ }\href {\doibase
  10.1103/PhysRevD.101.036006} {\bibfield  {journal} {\bibinfo  {journal}
  {Phys. Rev. D}\ }\textbf {\bibinfo {volume} {101}},\ \bibinfo {pages}
  {036006} (\bibinfo {year} {2020})},\ \bibinfo {note} {[Erratum: Phys.Rev.D
  102, 019904 (2020)]},\ \Eprint {http://arxiv.org/abs/1911.03482}
  {arXiv:1911.03482 [hep-ph]} \BibitemShut {NoStop}%
\bibitem [{\citenamefont {Migdal}(1939)}]{Migdal1939}%
  \BibitemOpen
  \bibfield  {author} {\bibinfo {author} {\bibfnamefont {A.}~\bibnamefont
  {Migdal}},\ }\href@noop {} {\bibfield  {journal} {\bibinfo  {journal} {Sov.
  Phys. JETP}\ }\textbf {\bibinfo {volume} {9}},\ \bibinfo {pages} {1163}
  (\bibinfo {year} {1939})}\BibitemShut {NoStop}%
\bibitem [{\citenamefont {Vergados}\ and\ \citenamefont
  {Ejiri}(2005)}]{Vergados:2004bm}%
  \BibitemOpen
  \bibfield  {author} {\bibinfo {author} {\bibfnamefont {J.~D.}\ \bibnamefont
  {Vergados}}\ and\ \bibinfo {author} {\bibfnamefont {H.}~\bibnamefont
  {Ejiri}},\ }\href {\doibase 10.1016/j.physletb.2004.11.085} {\bibfield
  {journal} {\bibinfo  {journal} {Phys. Lett.}\ }\textbf {\bibinfo {volume}
  {B606}},\ \bibinfo {pages} {313} (\bibinfo {year} {2005})},\ \Eprint
  {http://arxiv.org/abs/hep-ph/0401151} {arXiv:hep-ph/0401151 [hep-ph]}
  \BibitemShut {NoStop}%
%%CITATION = HEP-PH/0401151;%%
\bibitem [{\citenamefont {Moustakidis}\ \emph {et~al.}(2005)\citenamefont
  {Moustakidis}, \citenamefont {Vergados},\ and\ \citenamefont
  {Ejiri}}]{Moustakidis:2005gx}%
  \BibitemOpen
  \bibfield  {author} {\bibinfo {author} {\bibfnamefont {C.~C.}\ \bibnamefont
  {Moustakidis}}, \bibinfo {author} {\bibfnamefont {J.~D.}\ \bibnamefont
  {Vergados}}, \ and\ \bibinfo {author} {\bibfnamefont {H.}~\bibnamefont
  {Ejiri}},\ }\href {\doibase 10.1016/j.nuclphysb.2005.08.033} {\bibfield
  {journal} {\bibinfo  {journal} {Nucl. Phys.}\ }\textbf {\bibinfo {volume}
  {B727}},\ \bibinfo {pages} {406} (\bibinfo {year} {2005})},\ \Eprint
  {http://arxiv.org/abs/hep-ph/0507123} {arXiv:hep-ph/0507123 [hep-ph]}
  \BibitemShut {NoStop}%
%%CITATION = HEP-PH/0507123;%%
\bibitem [{\citenamefont {Bernabei}\ \emph {et~al.}(2007)\citenamefont
  {Bernabei} \emph {et~al.}}]{Bernabei:2007jz}%
  \BibitemOpen
  \bibfield  {author} {\bibinfo {author} {\bibfnamefont {R.}~\bibnamefont
  {Bernabei}} \emph {et~al.},\ }\href {\doibase 10.1142/S0217751X07037093}
  {\bibfield  {journal} {\bibinfo  {journal} {Int. J. Mod. Phys.}\ }\textbf
  {\bibinfo {volume} {A22}},\ \bibinfo {pages} {3155} (\bibinfo {year}
  {2007})},\ \Eprint {http://arxiv.org/abs/0706.1421} {arXiv:0706.1421
  [astro-ph]} \BibitemShut {NoStop}%
%%CITATION = ARXIV:0706.1421;%%
\bibitem [{\citenamefont {Ibe}\ \emph {et~al.}(2018)\citenamefont {Ibe},
  \citenamefont {Nakano}, \citenamefont {Shoji},\ and\ \citenamefont
  {Suzuki}}]{Ibe:2017yqa}%
  \BibitemOpen
  \bibfield  {author} {\bibinfo {author} {\bibfnamefont {M.}~\bibnamefont
  {Ibe}}, \bibinfo {author} {\bibfnamefont {W.}~\bibnamefont {Nakano}},
  \bibinfo {author} {\bibfnamefont {Y.}~\bibnamefont {Shoji}}, \ and\ \bibinfo
  {author} {\bibfnamefont {K.}~\bibnamefont {Suzuki}},\ }\href {\doibase
  10.1007/JHEP03(2018)194} {\bibfield  {journal} {\bibinfo  {journal} {JHEP}\
  }\textbf {\bibinfo {volume} {03}},\ \bibinfo {pages} {194} (\bibinfo {year}
  {2018})},\ \Eprint {http://arxiv.org/abs/1707.07258} {arXiv:1707.07258
  [hep-ph]} \BibitemShut {NoStop}%
%%CITATION = ARXIV:1707.07258;%%
\bibitem [{\citenamefont {Dolan}\ \emph {et~al.}(2018)\citenamefont {Dolan},
  \citenamefont {Kahlhoefer},\ and\ \citenamefont {McCabe}}]{Dolan:2017xbu}%
  \BibitemOpen
  \bibfield  {author} {\bibinfo {author} {\bibfnamefont {M.~J.}\ \bibnamefont
  {Dolan}}, \bibinfo {author} {\bibfnamefont {F.}~\bibnamefont {Kahlhoefer}}, \
  and\ \bibinfo {author} {\bibfnamefont {C.}~\bibnamefont {McCabe}},\ }\href
  {\doibase 10.1103/PhysRevLett.121.101801} {\bibfield  {journal} {\bibinfo
  {journal} {Phys. Rev. Lett.}\ }\textbf {\bibinfo {volume} {121}},\ \bibinfo
  {pages} {101801} (\bibinfo {year} {2018})},\ \Eprint
  {http://arxiv.org/abs/1711.09906} {arXiv:1711.09906 [hep-ph]} \BibitemShut
  {NoStop}%
%%CITATION = ARXIV:1711.09906;%%
\bibitem [{\citenamefont {Bell}\ \emph {et~al.}(2019)\citenamefont {Bell},
  \citenamefont {Dent}, \citenamefont {Newstead}, \citenamefont {Sabharwale},\
  and\ \citenamefont {Weiler}}]{Bell:2019egg}%
  \BibitemOpen
  \bibfield  {author} {\bibinfo {author} {\bibfnamefont {N.~F.}\ \bibnamefont
  {Bell}}, \bibinfo {author} {\bibfnamefont {J.~B.}\ \bibnamefont {Dent}},
  \bibinfo {author} {\bibfnamefont {J.~L.}\ \bibnamefont {Newstead}}, \bibinfo
  {author} {\bibfnamefont {S.}~\bibnamefont {Sabharwale}}, \ and\ \bibinfo
  {author} {\bibfnamefont {T.~J.}\ \bibnamefont {Weiler}},\ }\href@noop {} {\
  (\bibinfo {year} {2019})},\ \Eprint {http://arxiv.org/abs/1905.00046}
  {arXiv:1905.00046 [hep-ph]} \BibitemShut {NoStop}%
%%CITATION = ARXIV:1905.00046;%%
\bibitem [{\citenamefont {Baxter}\ \emph {et~al.}(2020)\citenamefont {Baxter},
  \citenamefont {Kahn},\ and\ \citenamefont {Krnjaic}}]{Baxter:2019pnz}%
  \BibitemOpen
  \bibfield  {author} {\bibinfo {author} {\bibfnamefont {D.}~\bibnamefont
  {Baxter}}, \bibinfo {author} {\bibfnamefont {Y.}~\bibnamefont {Kahn}}, \ and\
  \bibinfo {author} {\bibfnamefont {G.}~\bibnamefont {Krnjaic}},\ }\href
  {\doibase 10.1103/PhysRevD.101.076014} {\bibfield  {journal} {\bibinfo
  {journal} {Phys. Rev. D}\ }\textbf {\bibinfo {volume} {101}},\ \bibinfo
  {pages} {076014} (\bibinfo {year} {2020})},\ \Eprint
  {http://arxiv.org/abs/1908.00012} {arXiv:1908.00012 [hep-ph]} \BibitemShut
  {NoStop}%
\bibitem [{\citenamefont {Essig}\ \emph {et~al.}(2020)\citenamefont {Essig},
  \citenamefont {Pradler}, \citenamefont {Sholapurkar},\ and\ \citenamefont
  {Yu}}]{Essig:2019xkx}%
  \BibitemOpen
  \bibfield  {author} {\bibinfo {author} {\bibfnamefont {R.}~\bibnamefont
  {Essig}}, \bibinfo {author} {\bibfnamefont {J.}~\bibnamefont {Pradler}},
  \bibinfo {author} {\bibfnamefont {M.}~\bibnamefont {Sholapurkar}}, \ and\
  \bibinfo {author} {\bibfnamefont {T.-T.}\ \bibnamefont {Yu}},\ }\href
  {\doibase 10.1103/PhysRevLett.124.021801} {\bibfield  {journal} {\bibinfo
  {journal} {Phys. Rev. Lett.}\ }\textbf {\bibinfo {volume} {124}},\ \bibinfo
  {pages} {021801} (\bibinfo {year} {2020})},\ \Eprint
  {http://arxiv.org/abs/1908.10881} {arXiv:1908.10881 [hep-ph]} \BibitemShut
  {NoStop}%
\bibitem [{\citenamefont {Essig}\ \emph {et~al.}(2016)\citenamefont {Essig},
  \citenamefont {Fernandez-Serra}, \citenamefont {Mardon}, \citenamefont
  {Soto}, \citenamefont {Volansky},\ and\ \citenamefont {Yu}}]{Essig:2015cda}%
  \BibitemOpen
  \bibfield  {author} {\bibinfo {author} {\bibfnamefont {R.}~\bibnamefont
  {Essig}}, \bibinfo {author} {\bibfnamefont {M.}~\bibnamefont
  {Fernandez-Serra}}, \bibinfo {author} {\bibfnamefont {J.}~\bibnamefont
  {Mardon}}, \bibinfo {author} {\bibfnamefont {A.}~\bibnamefont {Soto}},
  \bibinfo {author} {\bibfnamefont {T.}~\bibnamefont {Volansky}}, \ and\
  \bibinfo {author} {\bibfnamefont {T.-T.}\ \bibnamefont {Yu}},\ }\href
  {\doibase 10.1007/JHEP05(2016)046} {\bibfield  {journal} {\bibinfo  {journal}
  {JHEP}\ }\textbf {\bibinfo {volume} {05}},\ \bibinfo {pages} {046} (\bibinfo
  {year} {2016})},\ \Eprint {http://arxiv.org/abs/1509.01598} {arXiv:1509.01598
  [hep-ph]} \BibitemShut {NoStop}%
\bibitem [{\citenamefont {de~Salas}(2020)}]{deSalas:2019rdi}%
  \BibitemOpen
  \bibfield  {author} {\bibinfo {author} {\bibfnamefont {P.~F.}\ \bibnamefont
  {de~Salas}},\ }\href {\doibase 10.1088/1742-6596/1468/1/012020} {\bibfield
  {journal} {\bibinfo  {journal} {J. Phys. Conf. Ser.}\ }\textbf {\bibinfo
  {volume} {1468}},\ \bibinfo {pages} {012020} (\bibinfo {year} {2020})},\
  \Eprint {http://arxiv.org/abs/1910.14366} {arXiv:1910.14366 [astro-ph.GA]}
  \BibitemShut {NoStop}%
\bibitem [{\citenamefont {Schober}(2014)}]{schober2014introduction}%
  \BibitemOpen
  \bibfield  {author} {\bibinfo {author} {\bibfnamefont {H.}~\bibnamefont
  {Schober}},\ }\href@noop {} {\bibfield  {journal} {\bibinfo  {journal}
  {Journal of Neutron Research}\ }\textbf {\bibinfo {volume} {17}},\ \bibinfo
  {pages} {109} (\bibinfo {year} {2014})}\BibitemShut {NoStop}%
\bibitem [{\citenamefont {deNiverville}\ \emph {et~al.}(2011)\citenamefont
  {deNiverville}, \citenamefont {Pospelov},\ and\ \citenamefont
  {Ritz}}]{deNiverville_2011}%
  \BibitemOpen
  \bibfield  {author} {\bibinfo {author} {\bibfnamefont {P.}~\bibnamefont
  {deNiverville}}, \bibinfo {author} {\bibfnamefont {M.}~\bibnamefont
  {Pospelov}}, \ and\ \bibinfo {author} {\bibfnamefont {A.}~\bibnamefont
  {Ritz}},\ }\href {\doibase 10.1103/physrevd.84.075020} {\bibfield  {journal}
  {\bibinfo  {journal} {Physical Review D}\ }\textbf {\bibinfo {volume} {84}}
  (\bibinfo {year} {2011}),\ 10.1103/physrevd.84.075020}\BibitemShut {NoStop}%
\bibitem [{\citenamefont {Banerjee}\ \emph {et~al.}(2019)\citenamefont
  {Banerjee} \emph {et~al.}}]{NA64:2019imj}%
  \BibitemOpen
  \bibfield  {author} {\bibinfo {author} {\bibfnamefont {D.}~\bibnamefont
  {Banerjee}} \emph {et~al.},\ }\href {\doibase 10.1103/PhysRevLett.123.121801}
  {\bibfield  {journal} {\bibinfo  {journal} {Phys. Rev. Lett.}\ }\textbf
  {\bibinfo {volume} {123}},\ \bibinfo {pages} {121801} (\bibinfo {year}
  {2019})},\ \Eprint {http://arxiv.org/abs/1906.00176} {arXiv:1906.00176
  [hep-ex]} \BibitemShut {NoStop}%
\bibitem [{\citenamefont {Batell}\ \emph {et~al.}(2014)\citenamefont {Batell},
  \citenamefont {Essig},\ and\ \citenamefont {Surujon}}]{Batell:2014mga}%
  \BibitemOpen
  \bibfield  {author} {\bibinfo {author} {\bibfnamefont {B.}~\bibnamefont
  {Batell}}, \bibinfo {author} {\bibfnamefont {R.}~\bibnamefont {Essig}}, \
  and\ \bibinfo {author} {\bibfnamefont {Z.}~\bibnamefont {Surujon}},\ }\href
  {\doibase 10.1103/PhysRevLett.113.171802} {\bibfield  {journal} {\bibinfo
  {journal} {Phys. Rev. Lett.}\ }\textbf {\bibinfo {volume} {113}},\ \bibinfo
  {pages} {171802} (\bibinfo {year} {2014})},\ \Eprint
  {http://arxiv.org/abs/1406.2698} {arXiv:1406.2698 [hep-ph]} \BibitemShut
  {NoStop}%
\bibitem [{\citenamefont {Marsicano}\ \emph {et~al.}(2018)\citenamefont
  {Marsicano}, \citenamefont {Battaglieri}, \citenamefont {Bond\'{\i}},
  \citenamefont {Carvajal}, \citenamefont {Celentano}, \citenamefont
  {De~Napoli}, \citenamefont {De~Vita}, \citenamefont {Nardi}, \citenamefont
  {Raggi},\ and\ \citenamefont {Valente}}]{PhysRevLett.121.041802}%
  \BibitemOpen
  \bibfield  {author} {\bibinfo {author} {\bibfnamefont {L.}~\bibnamefont
  {Marsicano}}, \bibinfo {author} {\bibfnamefont {M.}~\bibnamefont
  {Battaglieri}}, \bibinfo {author} {\bibfnamefont {M.}~\bibnamefont
  {Bond\'{\i}}}, \bibinfo {author} {\bibfnamefont {C.~D.~R.}\ \bibnamefont
  {Carvajal}}, \bibinfo {author} {\bibfnamefont {A.}~\bibnamefont {Celentano}},
  \bibinfo {author} {\bibfnamefont {M.}~\bibnamefont {De~Napoli}}, \bibinfo
  {author} {\bibfnamefont {R.}~\bibnamefont {De~Vita}}, \bibinfo {author}
  {\bibfnamefont {E.}~\bibnamefont {Nardi}}, \bibinfo {author} {\bibfnamefont
  {M.}~\bibnamefont {Raggi}}, \ and\ \bibinfo {author} {\bibfnamefont
  {P.}~\bibnamefont {Valente}},\ }\href {\doibase
  10.1103/PhysRevLett.121.041802} {\bibfield  {journal} {\bibinfo  {journal}
  {Phys. Rev. Lett.}\ }\textbf {\bibinfo {volume} {121}},\ \bibinfo {pages}
  {041802} (\bibinfo {year} {2018})}\BibitemShut {NoStop}%
\bibitem [{\citenamefont {Izaguirre}\ \emph {et~al.}(2013)\citenamefont
  {Izaguirre}, \citenamefont {Krnjaic}, \citenamefont {Schuster},\ and\
  \citenamefont {Toro}}]{Izaguirre:2013uxa}%
  \BibitemOpen
  \bibfield  {author} {\bibinfo {author} {\bibfnamefont {E.}~\bibnamefont
  {Izaguirre}}, \bibinfo {author} {\bibfnamefont {G.}~\bibnamefont {Krnjaic}},
  \bibinfo {author} {\bibfnamefont {P.}~\bibnamefont {Schuster}}, \ and\
  \bibinfo {author} {\bibfnamefont {N.}~\bibnamefont {Toro}},\ }\href {\doibase
  10.1103/PhysRevD.88.114015} {\bibfield  {journal} {\bibinfo  {journal} {Phys.
  Rev. D}\ }\textbf {\bibinfo {volume} {88}},\ \bibinfo {pages} {114015}
  (\bibinfo {year} {2013})},\ \Eprint {http://arxiv.org/abs/1307.6554}
  {arXiv:1307.6554 [hep-ph]} \BibitemShut {NoStop}%
\bibitem [{\citenamefont {Essig}\ \emph {et~al.}(2013)\citenamefont {Essig},
  \citenamefont {Mardon}, \citenamefont {Papucci}, \citenamefont {Volansky},\
  and\ \citenamefont {Zhong}}]{Essig:2013vha}%
  \BibitemOpen
  \bibfield  {author} {\bibinfo {author} {\bibfnamefont {R.}~\bibnamefont
  {Essig}}, \bibinfo {author} {\bibfnamefont {J.}~\bibnamefont {Mardon}},
  \bibinfo {author} {\bibfnamefont {M.}~\bibnamefont {Papucci}}, \bibinfo
  {author} {\bibfnamefont {T.}~\bibnamefont {Volansky}}, \ and\ \bibinfo
  {author} {\bibfnamefont {Y.-M.}\ \bibnamefont {Zhong}},\ }\href {\doibase
  10.1007/JHEP11(2013)167} {\bibfield  {journal} {\bibinfo  {journal} {JHEP}\
  }\textbf {\bibinfo {volume} {11}},\ \bibinfo {pages} {167} (\bibinfo {year}
  {2013})},\ \Eprint {http://arxiv.org/abs/1309.5084} {arXiv:1309.5084
  [hep-ph]} \BibitemShut {NoStop}%
\bibitem [{\citenamefont {Lees}\ \emph {et~al.}(2017)\citenamefont {Lees} \emph
  {et~al.}}]{Lees:2017lec}%
  \BibitemOpen
  \bibfield  {author} {\bibinfo {author} {\bibfnamefont {J.}~\bibnamefont
  {Lees}} \emph {et~al.} (\bibinfo {collaboration} {BaBar}),\ }\href {\doibase
  10.1103/PhysRevLett.119.131804} {\bibfield  {journal} {\bibinfo  {journal}
  {Phys. Rev. Lett.}\ }\textbf {\bibinfo {volume} {119}},\ \bibinfo {pages}
  {131804} (\bibinfo {year} {2017})},\ \Eprint
  {http://arxiv.org/abs/1702.03327} {arXiv:1702.03327 [hep-ex]} \BibitemShut
  {NoStop}%
\bibitem [{\citenamefont {Aguilar-Arevalo}(2020)}]{Aguilar_Arevalo_2020}%
  \BibitemOpen
  \bibfield  {author} {\bibinfo {author} {\bibfnamefont {A.~A.}\ \bibnamefont
  {Aguilar-Arevalo}},\ }\href {\doibase 10.1088/1742-6596/1342/1/012055}
  {\bibfield  {journal} {\bibinfo  {journal} {Journal of Physics: Conference
  Series}\ }\textbf {\bibinfo {volume} {1342}},\ \bibinfo {pages} {012055}
  (\bibinfo {year} {2020})}\BibitemShut {NoStop}%
\bibitem [{\citenamefont {Aprile}\ \emph {et~al.}(2019)\citenamefont {Aprile}
  \emph {et~al.}}]{Aprile:2019xxb}%
  \BibitemOpen
  \bibfield  {author} {\bibinfo {author} {\bibfnamefont {E.}~\bibnamefont
  {Aprile}} \emph {et~al.} (\bibinfo {collaboration} {XENON}),\ }\href
  {\doibase 10.1103/PhysRevLett.123.251801} {\bibfield  {journal} {\bibinfo
  {journal} {Phys. Rev. Lett.}\ }\textbf {\bibinfo {volume} {123}},\ \bibinfo
  {pages} {251801} (\bibinfo {year} {2019})},\ \Eprint
  {http://arxiv.org/abs/1907.11485} {arXiv:1907.11485 [hep-ex]} \BibitemShut
  {NoStop}%
\bibitem [{\citenamefont {Abdelhameed}\ \emph
  {et~al.}(2019{\natexlab{a}})\citenamefont {Abdelhameed} \emph
  {et~al.}}]{PhysRevD.100.102002}%
  \BibitemOpen
  \bibfield  {author} {\bibinfo {author} {\bibfnamefont {A.}~\bibnamefont
  {Abdelhameed}} \emph {et~al.} (\bibinfo {collaboration} {CRESST
  Collaboration}),\ }\href {\doibase 10.1103/PhysRevD.100.102002} {\bibfield
  {journal} {\bibinfo  {journal} {Phys. Rev. D}\ }\textbf {\bibinfo {volume}
  {100}},\ \bibinfo {pages} {102002} (\bibinfo {year}
  {2019}{\natexlab{a}})}\BibitemShut {NoStop}%
\bibitem [{\citenamefont {Abdelhameed}\ \emph
  {et~al.}(2019{\natexlab{b}})\citenamefont {Abdelhameed} \emph
  {et~al.}}]{Abdelhameed:2019hmk}%
  \BibitemOpen
  \bibfield  {author} {\bibinfo {author} {\bibfnamefont {A.}~\bibnamefont
  {Abdelhameed}} \emph {et~al.} (\bibinfo {collaboration} {CRESST}),\ }\href
  {\doibase 10.1103/PhysRevD.100.102002} {\bibfield  {journal} {\bibinfo
  {journal} {Phys. Rev. D}\ }\textbf {\bibinfo {volume} {100}},\ \bibinfo
  {pages} {102002} (\bibinfo {year} {2019}{\natexlab{b}})},\ \Eprint
  {http://arxiv.org/abs/1904.00498} {arXiv:1904.00498 [astro-ph.CO]}
  \BibitemShut {NoStop}%
\bibitem [{\citenamefont {Izaguirre}\ \emph {et~al.}(2015)\citenamefont
  {Izaguirre}, \citenamefont {Krnjaic}, \citenamefont {Schuster},\ and\
  \citenamefont {Toro}}]{Izaguirre:2015yja}%
  \BibitemOpen
  \bibfield  {author} {\bibinfo {author} {\bibfnamefont {E.}~\bibnamefont
  {Izaguirre}}, \bibinfo {author} {\bibfnamefont {G.}~\bibnamefont {Krnjaic}},
  \bibinfo {author} {\bibfnamefont {P.}~\bibnamefont {Schuster}}, \ and\
  \bibinfo {author} {\bibfnamefont {N.}~\bibnamefont {Toro}},\ }\href {\doibase
  10.1103/PhysRevLett.115.251301} {\bibfield  {journal} {\bibinfo  {journal}
  {Phys. Rev. Lett.}\ }\textbf {\bibinfo {volume} {115}},\ \bibinfo {pages}
  {251301} (\bibinfo {year} {2015})},\ \Eprint
  {http://arxiv.org/abs/1505.00011} {arXiv:1505.00011 [hep-ph]} \BibitemShut
  {NoStop}%
\bibitem [{\citenamefont {Berlin}\ \emph {et~al.}(2019)\citenamefont {Berlin},
  \citenamefont {Blinov}, \citenamefont {Krnjaic}, \citenamefont {Schuster},\
  and\ \citenamefont {Toro}}]{Berlin_2019}%
  \BibitemOpen
  \bibfield  {author} {\bibinfo {author} {\bibfnamefont {A.}~\bibnamefont
  {Berlin}}, \bibinfo {author} {\bibfnamefont {N.}~\bibnamefont {Blinov}},
  \bibinfo {author} {\bibfnamefont {G.}~\bibnamefont {Krnjaic}}, \bibinfo
  {author} {\bibfnamefont {P.}~\bibnamefont {Schuster}}, \ and\ \bibinfo
  {author} {\bibfnamefont {N.}~\bibnamefont {Toro}},\ }\href {\doibase
  10.1103/physrevd.99.075001} {\bibfield  {journal} {\bibinfo  {journal}
  {Physical Review D}\ }\textbf {\bibinfo {volume} {99}} (\bibinfo {year}
  {2019}),\ 10.1103/physrevd.99.075001}\BibitemShut {NoStop}%
\bibitem [{\citenamefont {Schiff}(1951)}]{Schiff:1951zza}%
  \BibitemOpen
  \bibfield  {author} {\bibinfo {author} {\bibfnamefont {L.}~\bibnamefont
  {Schiff}},\ }\href {\doibase 10.1103/PhysRev.83.252} {\bibfield  {journal}
  {\bibinfo  {journal} {Phys. Rev.}\ }\textbf {\bibinfo {volume} {83}},\
  \bibinfo {pages} {252} (\bibinfo {year} {1951})}\BibitemShut {NoStop}%
\bibitem [{\citenamefont {Tsai}(1974)}]{Tsai:1973py}%
  \BibitemOpen
  \bibfield  {author} {\bibinfo {author} {\bibfnamefont {Y.-S.}\ \bibnamefont
  {Tsai}},\ }\href {\doibase 10.1103/RevModPhys.46.815} {\bibfield  {journal}
  {\bibinfo  {journal} {Rev. Mod. Phys.}\ }\textbf {\bibinfo {volume} {46}},\
  \bibinfo {pages} {815} (\bibinfo {year} {1974})},\ \bibinfo {note} {[Erratum:
  Rev.Mod.Phys. 49, 421--423 (1977)]}\BibitemShut {NoStop}%
\bibitem [{\citenamefont {Emken}\ \emph {et~al.}(2019)\citenamefont {Emken},
  \citenamefont {Essig}, \citenamefont {Kouvaris},\ and\ \citenamefont
  {Sholapurkar}}]{Emken:2019tni}%
  \BibitemOpen
  \bibfield  {author} {\bibinfo {author} {\bibfnamefont {T.}~\bibnamefont
  {Emken}}, \bibinfo {author} {\bibfnamefont {R.}~\bibnamefont {Essig}},
  \bibinfo {author} {\bibfnamefont {C.}~\bibnamefont {Kouvaris}}, \ and\
  \bibinfo {author} {\bibfnamefont {M.}~\bibnamefont {Sholapurkar}},\ }\href
  {\doibase 10.1088/1475-7516/2019/09/070} {\bibfield  {journal} {\bibinfo
  {journal} {JCAP}\ }\textbf {\bibinfo {volume} {09}},\ \bibinfo {pages} {070}
  (\bibinfo {year} {2019})},\ \Eprint {http://arxiv.org/abs/1905.06348}
  {arXiv:1905.06348 [hep-ph]} \BibitemShut {NoStop}%
\bibitem [{\citenamefont {Aghanim}\ \emph {et~al.}(2020)\citenamefont {Aghanim}
  \emph {et~al.}}]{Aghanim:2018eyx}%
  \BibitemOpen
  \bibfield  {author} {\bibinfo {author} {\bibfnamefont {N.}~\bibnamefont
  {Aghanim}} \emph {et~al.} (\bibinfo {collaboration} {Planck}),\ }\href
  {\doibase 10.1051/0004-6361/201833910} {\bibfield  {journal} {\bibinfo
  {journal} {Astron. Astrophys.}\ }\textbf {\bibinfo {volume} {641}},\ \bibinfo
  {pages} {A6} (\bibinfo {year} {2020})},\ \Eprint
  {http://arxiv.org/abs/1807.06209} {arXiv:1807.06209 [astro-ph.CO]}
  \BibitemShut {NoStop}%
\bibitem [{\citenamefont {Kurinsky}\ \emph {et~al.}(2020)\citenamefont
  {Kurinsky}, \citenamefont {Baxter}, \citenamefont {Kahn},\ and\ \citenamefont
  {Krnjaic}}]{Kurinsky:2020dpb}%
  \BibitemOpen
  \bibfield  {author} {\bibinfo {author} {\bibfnamefont {N.}~\bibnamefont
  {Kurinsky}}, \bibinfo {author} {\bibfnamefont {D.}~\bibnamefont {Baxter}},
  \bibinfo {author} {\bibfnamefont {Y.}~\bibnamefont {Kahn}}, \ and\ \bibinfo
  {author} {\bibfnamefont {G.}~\bibnamefont {Krnjaic}},\ }\href {\doibase
  10.1103/PhysRevD.102.015017} {\bibfield  {journal} {\bibinfo  {journal}
  {Phys. Rev. D}\ }\textbf {\bibinfo {volume} {102}},\ \bibinfo {pages}
  {015017} (\bibinfo {year} {2020})},\ \Eprint
  {http://arxiv.org/abs/2002.06937} {arXiv:2002.06937 [hep-ph]} \BibitemShut
  {NoStop}%
\bibitem [{\citenamefont {Kozaczuk}\ and\ \citenamefont
  {Lin}(2020)}]{Kozaczuk:2020uzb}%
  \BibitemOpen
  \bibfield  {author} {\bibinfo {author} {\bibfnamefont {J.}~\bibnamefont
  {Kozaczuk}}\ and\ \bibinfo {author} {\bibfnamefont {T.}~\bibnamefont {Lin}},\
  }\href {\doibase 10.1103/PhysRevD.101.123012} {\bibfield  {journal} {\bibinfo
   {journal} {Phys. Rev. D}\ }\textbf {\bibinfo {volume} {101}},\ \bibinfo
  {pages} {123012} (\bibinfo {year} {2020})},\ \Eprint
  {http://arxiv.org/abs/2003.12077} {arXiv:2003.12077 [hep-ph]} \BibitemShut
  {NoStop}%
\bibitem [{\citenamefont {Essig}\ \emph {et~al.}(2012)\citenamefont {Essig},
  \citenamefont {Mardon},\ and\ \citenamefont {Volansky}}]{Essig:2011nj}%
  \BibitemOpen
  \bibfield  {author} {\bibinfo {author} {\bibfnamefont {R.}~\bibnamefont
  {Essig}}, \bibinfo {author} {\bibfnamefont {J.}~\bibnamefont {Mardon}}, \
  and\ \bibinfo {author} {\bibfnamefont {T.}~\bibnamefont {Volansky}},\ }\href
  {\doibase 10.1103/PhysRevD.85.076007} {\bibfield  {journal} {\bibinfo
  {journal} {Phys. Rev. D}\ }\textbf {\bibinfo {volume} {85}},\ \bibinfo
  {pages} {076007} (\bibinfo {year} {2012})},\ \Eprint
  {http://arxiv.org/abs/1108.5383} {arXiv:1108.5383 [hep-ph]} \BibitemShut
  {NoStop}%
\bibitem [{\citenamefont {Kurinsky}\ \emph {et~al.}(2017)\citenamefont
  {Kurinsky}, \citenamefont {Brink}, \citenamefont {Partridge}, \citenamefont
  {Cabrera},\ and\ \citenamefont {Pyle}}]{Kurinsky:2016fvj}%
  \BibitemOpen
  \bibfield  {author} {\bibinfo {author} {\bibfnamefont {N.}~\bibnamefont
  {Kurinsky}}, \bibinfo {author} {\bibfnamefont {P.}~\bibnamefont {Brink}},
  \bibinfo {author} {\bibfnamefont {R.}~\bibnamefont {Partridge}}, \bibinfo
  {author} {\bibfnamefont {B.}~\bibnamefont {Cabrera}}, \ and\ \bibinfo
  {author} {\bibfnamefont {M.}~\bibnamefont {Pyle}} (\bibinfo {collaboration}
  {SuperCDMS}),\ }\href {\doibase 10.22323/1.282.1116} {\bibfield  {journal}
  {\bibinfo  {journal} {PoS}\ }\textbf {\bibinfo {volume} {ICHEP2016}},\
  \bibinfo {pages} {1116} (\bibinfo {year} {2017})},\ \Eprint
  {http://arxiv.org/abs/1611.04083} {arXiv:1611.04083 [physics.ins-det]}
  \BibitemShut {NoStop}%
\bibitem [{\citenamefont {Agnese}\ \emph {et~al.}(2018)\citenamefont {Agnese}
  \emph {et~al.}}]{Agnese:2018col}%
  \BibitemOpen
  \bibfield  {author} {\bibinfo {author} {\bibfnamefont {R.}~\bibnamefont
  {Agnese}} \emph {et~al.} (\bibinfo {collaboration} {SuperCDMS}),\ }\href
  {\doibase 10.1103/PhysRevLett.121.051301} {\bibfield  {journal} {\bibinfo
  {journal} {Phys. Rev. Lett.}\ }\textbf {\bibinfo {volume} {121}},\ \bibinfo
  {pages} {051301} (\bibinfo {year} {2018})},\ \bibinfo {note} {[Erratum:
  Phys.Rev.Lett. 122, 069901 (2019)]},\ \Eprint
  {http://arxiv.org/abs/1804.10697} {arXiv:1804.10697 [hep-ex]} \BibitemShut
  {NoStop}%
\bibitem [{\citenamefont {Alkhatib}\ \emph {et~al.}(2020)\citenamefont
  {Alkhatib} \emph {et~al.}}]{Alkhatib:2020slm}%
  \BibitemOpen
  \bibfield  {author} {\bibinfo {author} {\bibfnamefont {I.}~\bibnamefont
  {Alkhatib}} \emph {et~al.} (\bibinfo {collaboration} {SuperCDMS}),\
  }\href@noop {} {\  (\bibinfo {year} {2020})},\ \Eprint
  {http://arxiv.org/abs/2007.14289} {arXiv:2007.14289 [hep-ex]} \BibitemShut
  {NoStop}%
\bibitem [{\citenamefont {Armengaud}\ \emph {et~al.}(2019)\citenamefont
  {Armengaud} \emph {et~al.}}]{Armengaud:2019kfj}%
  \BibitemOpen
  \bibfield  {author} {\bibinfo {author} {\bibfnamefont {E.}~\bibnamefont
  {Armengaud}} \emph {et~al.} (\bibinfo {collaboration} {EDELWEISS}),\ }\href
  {\doibase 10.1103/PhysRevD.99.082003} {\bibfield  {journal} {\bibinfo
  {journal} {Phys. Rev. D}\ }\textbf {\bibinfo {volume} {99}},\ \bibinfo
  {pages} {082003} (\bibinfo {year} {2019})},\ \Eprint
  {http://arxiv.org/abs/1901.03588} {arXiv:1901.03588 [astro-ph.GA]}
  \BibitemShut {NoStop}%
\bibitem [{\citenamefont {Arnaud}\ \emph {et~al.}(2020)\citenamefont {Arnaud}
  \emph {et~al.}}]{Arnaud:2020svb}%
  \BibitemOpen
  \bibfield  {author} {\bibinfo {author} {\bibfnamefont {Q.}~\bibnamefont
  {Arnaud}} \emph {et~al.} (\bibinfo {collaboration} {EDELWEISS}),\ }\href
  {\doibase 10.1103/PhysRevLett.125.141301} {\bibfield  {journal} {\bibinfo
  {journal} {Phys. Rev. Lett.}\ }\textbf {\bibinfo {volume} {125}},\ \bibinfo
  {pages} {141301} (\bibinfo {year} {2020})},\ \Eprint
  {http://arxiv.org/abs/2003.01046} {arXiv:2003.01046 [astro-ph.GA]}
  \BibitemShut {NoStop}%
\bibitem [{\citenamefont {Kurinsky}\ \emph {et~al.}(2019)\citenamefont
  {Kurinsky}, \citenamefont {Yu}, \citenamefont {Hochberg},\ and\ \citenamefont
  {Cabrera}}]{Kurinsky:2019pgb}%
  \BibitemOpen
  \bibfield  {author} {\bibinfo {author} {\bibfnamefont {N.~A.}\ \bibnamefont
  {Kurinsky}}, \bibinfo {author} {\bibfnamefont {T.~C.}\ \bibnamefont {Yu}},
  \bibinfo {author} {\bibfnamefont {Y.}~\bibnamefont {Hochberg}}, \ and\
  \bibinfo {author} {\bibfnamefont {B.}~\bibnamefont {Cabrera}},\ }\href
  {\doibase 10.1103/PhysRevD.99.123005} {\bibfield  {journal} {\bibinfo
  {journal} {Phys. Rev. D}\ }\textbf {\bibinfo {volume} {99}},\ \bibinfo
  {pages} {123005} (\bibinfo {year} {2019})},\ \Eprint
  {http://arxiv.org/abs/1901.07569} {arXiv:1901.07569 [hep-ex]} \BibitemShut
  {NoStop}%
\bibitem [{\citenamefont {Knapen}\ \emph {et~al.}(2020)\citenamefont {Knapen},
  \citenamefont {Kozaczuk},\ and\ \citenamefont {Lin}}]{toappear}%
  \BibitemOpen
  \bibfield  {author} {\bibinfo {author} {\bibfnamefont {S.}~\bibnamefont
  {Knapen}}, \bibinfo {author} {\bibfnamefont {J.}~\bibnamefont {Kozaczuk}}, \
  and\ \bibinfo {author} {\bibfnamefont {T.}~\bibnamefont {Lin}},\ }\href@noop
  {} {\bibfield  {journal} {\bibinfo  {journal} {``The Migdal effect in
  semiconductors"}\ } (\bibinfo {year} {2020})}\BibitemShut {NoStop}%
\end{thebibliography}%

\newpage

\onecolumngrid
\begin{center}
\textbf{\large 
 Dark Matter Direct Detection With Bound Nuclear Targets: The Poisson Phonon Tail
} \\ 
%\vspace{0.1in}
{ \it \large Supplementary Material}\\ 
%\vspace{0.05in}
{}
{Yonatan Kahn, Gordan Krnjaic, Bashi Mandava}

\end{center}
%%%%%%%%%% Merge with supplemental materials %%%%%%%%%%
\setcounter{equation}{0}
\setcounter{figure}{0}
\setcounter{table}{0}
\setcounter{section}{1}
\renewcommand{\theequation}{S\arabic{equation}}
\renewcommand{\thefigure}{S\arabic{figure}}
\renewcommand{\thetable}{S\arabic{table}}
\newcommand\ptwiddle[1]{\mathord{\mathop{#1}\limits^{\scriptscriptstyle(\sim)}}}

\section{1. Deriving The Scattering Rate}
The transition rate between nuclear states induced by DM-nuclear scattering can be computed using Fermi's Golden Rule. Following \cite{Essig:2015cda}, the scattering cross section times velocity
for the inelastic $\chi(\pp) N(0) \to \chi(\pp^\prime) N(\vec{n})$ transition
between individual oscillator levels $0\to (\vec{n})$ can be written in relativistic normalization as
\be
\label{sigmav}
\sigma v_{\{0\to\vec{n}\}}
 = \frac{1}{4 E_{\chi} E_{N}} 
 \int \frac{d^3 \qq}{(2\pi)^3}  \frac{1}{4 E'_{\chi} E'_{N}}
(2\pi) \delta(E_f - E_i)  \overline{| {\pazocal M}(q)|^2},
\ee
where primes denote final-state quantities. To calculate the momentum-space matrix element $\pazocal M(q)$, 
we begin by postulating a contact interaction between dark matter and nuclei of the form
\be
\label{VDM}
 \hat V(\mathbf{r}_\chi - \mathbf{r}_N ) =  \frac{{\pazocal M}_N}{4m_N m_\chi} \delta^3(\mathbf{r}_\chi-\mathbf{r}_N), 
\ee
where
$\rr_\chi$, $\rr_N$ are the DM and nuclear coordinates, respectively, and
 ${\pazocal M}_N$ is the dimensionless relativistic scattering matrix element between DM and a free nucleus, which is simply a constant for a contact interaction. For simplicity we consider the case of scattering mediated through a heavy dark photon, such that the fundamental interaction is between DM and protons; the matrix elements are related by
 \be
 {\pazocal M}_N = Z \frac{m_N}{m_p}  {\pazocal M}_p.
 \ee
 From this we can define the fiducial single
  proton cross section
\be
\overline \sigma_p \equiv \frac{     \mu_{\chi p}^2    |\overline{ {\pazocal M}_p|^2}      }{16 \pi m_\chi^2 m_p^2}~.
\ee
Using relativistic state normalization, the initial/final state wave functions in position space are 
\be
\label{psis}
|\pp;  0\rangle  &=& \sqrt{2m_\chi} \sqrt{2m_N} \,  \psi_0 (\mathbf{r}_N)  \, e^{i \mathbf{p}\cdot\mathbf{r}_\chi}  \\ 
|\pp^\prime; \vec{n}\rangle  &=& \sqrt{2m_\chi} \sqrt{2m_N}   \, \psi_{\vec{n}} (\mathbf{r}_N)   \, e^{i \mathbf{p}^\prime \cdot \mathbf{r}_\chi} ,
\ee
where we treat the dark matter as a free-particle plane wave and the $\psi_i(\rr_N)$ are harmonic
oscillator wavefunctions. The matrix element can then be written 
\be
 {\pazocal M}(q) = \langle \pp^\prime; \vec{n}  |  \hat V   |  \pp; 0  \rangle 
 &=& {\pazocal M}_N \int d^3 \rr_N \int d^3 \rr_\chi \, e^{i (\pp- \pp^\prime) \cdot \rr_\chi}  \psi^*_{\vec{n}}(\rr_N)   \delta^3(\rr_\chi - \rr_N)  \psi_0(\rr_N) 
 \nonumber \\
 &=& {\pazocal M}_N \int d^3 \rr_N  \psi^*_{\vec{n}}(\rr_N)  \, e^{i \qq \cdot \rr_N} \psi_0(\rr_N)
 \equiv {\pazocal M}_N \langle \vec{n} | \, e^{i \qq \cdot \hat{\rr}_N} |0\rangle~,
 \label{eq:MNf}
\ee
where $\qq \equiv \pp- \pp^\prime$ is the momentum transferred from the dark matter to the harmonic oscillator system.
The harmonic oscillator matrix element can be written in momentum space as follows:
  \be
\langle \vec{n} | \, e^{i \qq \cdot \hat \rr_N} |0\rangle&=&  
  \int d^3\mathbf r_N 
 \psi^*_{\vec{n}} (\mathbf{r}_N)  
 \, e^{i \qq \cdot \rr_N} 
 \psi_0(\mathbf{r}_N) 
 \nonumber \\
 &=& 
   \int d^3 \mathbf r_N 
  \int \frac{d^3 \mathbf p}{(2\pi)^3} 
  \int \frac{d^3 \mathbf k}{(2\pi)^3}  
  e^{i (\mathbf{q} + \mathbf{k}  - \mathbf{p} )\cdot \mathbf{r}_N}
\tilde  \phi^*_{\vec{n}} (\mathbf{p})  
\tilde  \phi_0(\mathbf{k})
  \nonumber \\
 &=& 
  \int \frac{d^3 \mathbf p}{(2\pi)^3}  
\tilde  \phi^*_{\vec{n}} (\mathbf{p})  
\tilde  \phi_0(\mathbf{p}- \mathbf{q})~,
\label{eq:matrixelementP}
  \ee
where $\tilde \phi_n$ are harmonic oscillator wave functions
in momentum space.
Note that, for comparison with the derivation in \cite{Essig:2015cda},
  here we adopt a normalization convention for which 
  \be
  \label{stupid}
  \int d^3 \mathbf r \,
 \psi^*_m (\mathbf{r})  
 \psi_n(\mathbf{r})
     =
  \int \frac{d^3 \mathbf q}{(2\pi)^3}   
\tilde  \phi^*_m (\mathbf{q})  
\tilde  \phi_n(\mathbf{q})   = \delta_{mn}~.
  \ee
  Putting it all together, we 
insert ${\pazocal M}(q)$ back into \Eq{sigmav} and replace $E_\chi, E_\chi' \to m_\chi$, $E_N, E'_N \to m_N$ in the non-relativistic limit. Squaring and summing over all oscillator levels such that $n_x + n_y + n_z = n$ yields the form factor $| f(\mathbf{q}, n)  |^2$ defined in \Eq{fnq}. Summing over all allowed energies (labeled by $n$) and scattering targets and integrating over the DM velocity distribution, we obtain the usual rate formula \cite{Essig:2015cda} 
stated in \Eq{eq:Rtot}:
  \be
  \label{eq:RtotSM}
  R =  N_T \frac{\rho_\chi}{m_\chi}  \frac{Z^2 \bar \sigma_p}{8\pi \mu_{\chi p}^2} \int \frac{d^3 \mathbf q}{q}  
  \sum_n\, | f(\mathbf{q}, n)  |^2   \eta(v_{\rm min}^{(n)})~,
  \ee
 where $\eta = \langle v^{-1} \theta(v-v_{\rm min} ) \rangle$ is the inverse mean speed and $v_{\rm min}$ 
 is the minimum DM velocity required to upscatter the nucleus into a state with energy $E_n = (n + 1/2)\omegaO$. Note that our derivation here is essentially identical to that of \cite{Essig:2015cda}, replacing DM scattering from bound electrons with DM scattering from bound nuclei. The only essential difference is the factor of $Z^2$ arising from coherent scattering over all protons in the nucleus.

\section{2. Evaluating The Matrix Element}

\subsection{A. Momentum Space Wavefunctions}

We begin by constructing the harmonic oscillator wavefunctions in momentum space, which is convenient for evaluating the matrix element where the argument of one wavefunction is translated by the momentum transfer $\qq$.
In momentum space, the position and momentum operators are represented as
$\hat{\mathbf{\pp}} \rightarrow \mathbf{p}$, 
$ \hat{\mathbf{\xx}} \rightarrow i \mathbf{\nabla}_p$,
so the time independent Schr\"{o}dinger equation with $\hbar = 1$ becomes
\be
\left(\frac{\mathbf{p}^{2}}{2 m_N}-\frac{m_N \omega_0^{2} }{2} \nabla^{2}\right) \phi(\mathbf{p})=
\sum_{i=x,y,z}\left(\frac{p_{i}^{2}}{2 m_N}-\frac{m_N \omega_0^{2} }{2} \frac{\partial^{2}}{\partial p_i^{2}}\right) \phi(\mathbf{p})=E \phi(\mathbf{p}) ~,~~
\ee
which we have written in terms of separable solutions $\phi(\mathbf{p})=\phi(p_x)\phi(p_y)\phi(p_z)$,
each satisfying $\hat H \phi_i= E_i \phi_i$, where $E=E_{1}+E_{2}+E_{3}$. Defining dimensionless quantities $\tilde{p} \equiv p/\sqrt{m_N\omegaO}$ and $\varepsilon\equiv E/\omegaO$, each eigenvalue equation becomes
\begin{equation}
    \phi''+(2\varepsilon-\tilde{p}^2)\phi=0\notag,
\end{equation}
where  $^\prime$ denotes differentiation with respect to $\tilde{p}$. 
For each $i=x,y,z$, the normalized solutions satisfy 
\be
\label{wavefunctions}
\phi_{n}\left(p_{i}\right)=A_{n} \exp \left(- \dfrac{p_{i}^{2}}{2  m_N \omegaO}\right) H_n  \! \left(\dfrac{p_{i}}{\sqrt{ m_N\omegaO}}\right)
~~,~~A_{n} = \dfrac{ A_0 }{ \sqrt{ 2^{n} n! }}~~,
\ee
where $H_n$ is an Hermite polynomial. In particular, the ground state has a Gaussian profile,
\be
\phi_{0}\left(p_{i}\right)=A_0\operatorname{exp}\left(-\dfrac{p_{i}^{2}}{2 m_N \omegaO }\right)   ~,~~A_0=\left( m_N\omegaO\pi \right) ^{-1/4}~.
\ee
with momentum spread of order $q_0 = \sqrt{2 m_N \omegaO}$ as claimed.
Note that the normalized wavefunctions~(\ref{wavefunctions}) satisfy the usual non-relativistic normalization convention
\be
\label{unit-convention}
\int d^3\pp\, \phi_m^*(\mathbf{p}) \phi_n(\mathbf{p}) = \delta_{mn},
\ee
which differs from the convention in \Eq{stupid} by $\tilde \phi_n = (2\pi)^{3/2} \phi_n$ which is more common in relativistic treatments. Throughout this paper
(including the remaining Supplementary Material), we use the convention in \Eq{unit-convention}.

\subsection{B. Poisson Distribution  }

According to \Eq{eq:matrixelementP}, we may write the matrix element between harmonic oscillator states in momentum space as
\be
\label{first-term}
 \langle \vec{n} |    e^{i \qq \cdot \mathbf{\hat r}_N} | 0 \rangle = \int   dp_xdp_ydp_z    \, \phi^*_n(\mathbf p)\phi_0(\mathbf p-\bf q)~. 
\ee
Since the wave functions are  separable in Cartesian coordinates, we need only evaluate the integral for a single component. For ease of notation, we will do this for the $x$-coordinate and write $n \equiv n_x$: 
\be
\label{1Dintegral}
\int_{-\infty}^\infty dp_x  \phi^*_n(p_x)\phi_0(p_x-q_x) 
&=&
A_0A_n  \int_{-\infty}^\infty dp_x \exp{\left(-\frac{p_x^2}{2m_N\omegaO}\right)}H_n \left(    \frac{p_x}{\sqrt{m_N \omegaO}} \right) \exp{\left(-\frac{(p_x-q_x)^2}{2m_N \omegaO }\right)} ~.
\ee
Defining the dimensionless variables 
$a\equiv p_x/\sqrt{m_N \omegaO}$ and $b \equiv q_x/\sqrt{m_N \omegaO}$, 
the right-hand side of \Eq{1Dintegral} is
\be
A_0A_n
\sqrt{m_N \omegaO}
\exp\left(-\frac{ b^2}{2} \right)
  \int_{-\infty}^\infty da 
  \exp \left(  - a^2 + a b \right)
H_n (a)~,
\ee
so our task reduces to evaluating the expression 
\be 
I_{n}(b) \equiv    \int_{-\infty}^\infty da \exp \left(-a^{2}+b a \right) H_{n}(a).
\ee
Note that the integrand is related to the generating functions for Hermite polynomials:
\be
    \exp \left(-t^{2}+2 a t\right)=\sum_{n=0}^{\infty} \frac{t^{n}}{n !} H_{n}(a) ~.
    \ee 
By performing a summation over all $I_n(b)$ as follows:
\be
\sum_{n=0}^{\infty} \frac{t^{n}}{n !}I_{n}(b) &=&\sum_{n=0}^{\infty} \frac{t^{n}}{n !} 
\int_{-\infty}^\infty
 da \exp \left(-a^{2}+b a\right) H_n(a)  
 \nonumber \\
&=&\int_{-\infty}^\infty da \operatorname{exp}\left(-a^{2}+b a-t^{2}+2 a t\right) 
%&=&  \exp \left(-t^{2}\right) \int_{-\infty}^\infty da  \exp \left[-a^{2}+a(2 t+b)\right] 
%\nonumber  \\
%&=&\sqrt{\pi} \exp \left(-t^{2}\right)  \exp \left[\frac{(b+2 t)^{2}}{4}\right]
%\nonumber  \\
=\sqrt{\pi} \exp \left(\frac{b^{2}}{4}\right) \exp (b t)~,
%\nonumber \\
%&=&\sqrt{\pi} \exp \left(\frac{b^{2}}{4}\right) 
%\sum_{n=0}^{\infty} \frac{t^{n}}{n !} b^{n},
\ee
and using the Taylor expansion of $\exp(bt)$ on the right hand side, we can read off the expression for individual $I_n(b)$ by matching terms of equal $n$ in the
summation:
\begin{equation}
\sum_{n=0}^{\infty} \frac{t^{n}}{n !}I_{n}(b) = \sqrt{\pi} \exp \left(\frac{b^{2}}{4}\right)  \sum_{n=0}^{\infty} \frac{t^{n}}{n !} b^n ~~\implies ~~
I_{n}(b) =\sqrt{\pi} \exp \left(\frac{b^{2}}{4}\right) b^{n} .
\end{equation}
Using this  result to evaluate our original Cartesian integral, we obtain (now restoring the index $n_x$)
\be
  \int_{-\infty}^\infty dp_x  \phi^*_{n_x}(p_x)\phi_0(p_x-q_x) 
%   &=&A_0A_{n_x}  \sqrt{m_N \omegaO}
%   \exp{ \left(   - \frac{b^2}{2}   \right)  }
%     I_n(b)\\
    &=&A_0A_{n_x}\sqrt{\pi m_N \omegaO} \operatorname{exp}\left(-\frac{q_x^2}{4m_N \omegaO}\right)\left(\frac{q_x}{\sqrt{m_N \omegaO}}\right)^{n_x}.
\ee
Since the integrals in \Eq{first-term} are identical in $p_x,p_y,p_z$, we can use
\Eq{wavefunctions} to write their product as 
\be
    \langle \vec{n} |   e^{     i \mathbf{\hat q} \cdot \mathbf{\hat r}_N }      | 0 \rangle
%    &=&A_0^3A_{n_x}A_{n_y}A_{n_z}(\pi m_N \omegaO)^{ 3/2 } \operatorname{exp}\left(\frac{-q^2}{4 m_N \omegaO}\right)  \frac{q_x^{n_x}q_y^{n_y}q_z^{n_z}}{{ (m_N \omegaO})^{    n/2    }   } 
%    \nonumber \\
    &=&
    \left[    2^n (n_x!n_y!n_z!)  \right]^{-1/2}
    \exp \left(\frac{-q^{2}}{4 m_N \omegaO }\right)\left(\frac{q}{\sqrt{m_N \omegaO}}\right)^{n}\left(\sin \theta \cos \phi)^{n_x}(\sin \theta \sin\phi )^{n_y}(\cos \theta \right)^{n_z},~~ ~~~~~~~
\ee
where $n = n_x + n_y + n_z$ and 
we have written the momentum components $q_i$ in spherical coordinates for future 
convenience. Squaring and summing over degenerate states, the form factor for exciting the $n^\text{th}$ energy level of the harmonic oscillator is given by
\begin{equation}
\label{eq:PN}
|f(n, \qq)|^2=\sum_{n_x + n_y + n_z = n} \left(\frac{1}{n_x!n_y!n_z!}\right) \exp \left(\frac{-q^{2}}{2 m_N \omegaO }\right)\left(\frac{q}{\sqrt{2m_N \omegaO }}\right)^{2n}\left(\sin \theta \cos \phi)^{2n_x}(\sin \theta \sin\phi )^{2n_y}(\cos \theta \right)^{2n_z},
\end{equation}
which resembles a Poisson distribution up to the angular factors. We now turn to 
performing an angular average of this expression to justify the full Poissonian form 
shown in \Eq{eq:fNq}.

\subsection{C. Angular Integrals}

Since phonon directions are not observable, we are interested in computing 
the angular average $\frac{1}{4\pi} \int d\Omega_{\qq} |f(n,\qq)|^2$, summing
over all occupation numbers that satisfy the constraint $n = n_x + n_y +n_x$. To average the angular part of \Eq{eq:PN}, we need to compute 
\begin{equation}
\begin{aligned}
\label{Fn}
F_{n}&=\frac{1}{4\pi} \sum_{n_{x}+n_{y}+n_{z}=n} \frac{1}{n_{x} ! n_{y} ! n_{z} !}
\int_{0}^{\pi} d \theta \int_{0}^{2 \pi} d \phi \, (\sin \theta)^{2 n_{x}+2 n_{y}+1}(\cos \theta)^{2 n_{z}}(\sin \phi)^{2 n_{y}}(\cos \phi)^{2 n_{x}}\\&=
\frac{1}{4\pi} \int_{0}^{\pi} d \theta \sin \theta \int_{0}^{2 \pi} d \phi \sum_{n_{x}+n_{y}+n_{z}=n} \frac{1}{n_{x} ! n_{y} ! n_{z} !}(\sin \theta\cos\phi)^{2 n_{x}}(\sin\theta\sin\phi)^{2 n_{y}}(\cos \theta)^{2 n_{z}}~.
\end{aligned}
\end{equation}
Let $(\sin \theta\cos\phi) ^2=A$, $(\sin\theta\sin\phi)^2=B$ and $(\cos \theta)^2=C$. The integrand with the summation is
\begin{equation}
\begin{aligned}
K_{n}&=\sum_{n_{x}+n_{y}+n_{z}=n} \frac{1}{n_{x} ! n_{y} ! n_{z} !}(\sin \theta\cos\phi)^{2 n_{x}}(\sin\theta\sin\phi)^{2 n_{y}}(\cos \theta)^{2 n_{z}}\\
&=\sum_{n_{x}+n_{y}+n_{z}=n} \frac{1}{n_{x} ! n_{y} ! n_{z} !}A^{ n_{x}}B^{ n_{y}}C^{ n_{z}}.
\end{aligned}
\end{equation}
Multiplying and dividing by $n!$, we have
\begin{equation}
K_{n}=\frac{1}{n!}\sum_{n_{x}+n_{y}+n_{z}=n} \frac{n! }{n_{x} ! n_{y} ! n_{z} !}A^{ n_{x}}B^{ n_{y}}C^{ n_{z}},
\end{equation}
where the summation is simply the multinomial expansion,
\begin{equation}
\sum_{k_{1}+k_{2}+\cdots+k_{m}=n}\frac{n !}{k_{1} ! k_{2} ! \cdots k_{m} !}x_{1}^{k_{1}} x_{2}^{k_{2}} \cdots x_{m}^{k_{m}}\\=\left(x_{1}+x_{2}+\cdots+x_{m}\right)^{n}.
\end{equation}
With this reorganization of terms, our integrand of interest simplifies considerably:
\begin{equation}
\begin{aligned}
K_{n} =\frac{1}{n!}\left(A+B+C\right)^{n}
=\frac{1}{n!}\left[(\sin \theta\cos\phi)^2+ (\sin\theta\sin\phi)^2+(\cos \theta)^2\right]^{n}
=\frac{1}{n!}~.
\end{aligned}
\end{equation}
Using this form, \Eq{Fn} becomes
\begin{equation}
F_{n}= \frac{1}{4\pi n!}\int_{0}^{\pi} d \theta \sin \theta  \int_{0}^{2 \pi} d \phi\ =\frac{1}{n! },
\end{equation}
as claimed in \Eq{eq:fNq}.

\subsection{D. Operator Algebra}
We can obtain the same result by using the algebra of creation and annihilation operators. The three-dimensional harmonic oscillator can be separated into three mutually-commuting sets of creation and annihilation operators,
\begin{equation}
[\ahat_i, \ahat^\dagger_i] = 1, \qquad i = 1, 2, 3,
\end{equation}
in terms of which the position and momentum operators $\hat{x}_i$ and $\hat{p}_i$ can be written
\begin{equation}
\hat{x}_i = \frac{1}{\sqrt{ 2 m_N \omegaO}} \left(\ahat_i + \ahat^\dagger_i \right), \qquad \hat{p}_i = i \sqrt{\frac{  m_N  \omegaO}{2}} \left(-\ahat_i + \ahat^\dagger_i\right).
\end{equation}
Since $\mathbf{q} \cdot \xhat_{N} = q_x \hat{x}_N + q_y \hat{y}_N + q_z \hat{z}_N$ and the spatial operators commute with each other, without loss of generality we can simply compute
\begin{equation}
|\langle 0 | e^{i q_x \hat{x}_N} | n_x \rangle |^2
\end{equation}
and copy the result for $n_y$ and $n_z$; multiplying these together gives the desired result for arbitrary $n$.

First let's write the exponential operator in terms of creation and annihilation operators:
\begin{equation}
\label{exp}
\exp( {i q_x \hat{x}_N})  = \exp\left[     \frac{i q_x}{\sqrt{2 m_N  \omega_0}} \left(\ahat + \ahat^\dagger \right) \right],
\end{equation}
where we have dropped subscript on $a$ for convenience. Using the Baker-Campbell-Hausdorff formula
for any operators $A$ and $B$, we can write 
\begin{equation}
\exp(A) \exp(B) = \exp\left(   A + B + \frac{1}{2}[A,B] + \frac{1}{12}[A,[A,B]] - \frac{1}{12}[B,[A,B]]\dots \right)~.
\end{equation}
Identifying $A  \to \ahat^\dagger$ and $B \to \ahat$, and noting that $[\ahat, \ahat^\dagger] = 1$ is a $c$-number, the
 series in \Eq{exp} truncates after the third term. Indeed, the third term is just a number, so we have the exact result
\begin{equation}
e^{\ahat^\dagger} e^{\ahat} = e^{\ahat + \ahat^\dagger - \frac{1}{2}}~,
\end{equation}
and since the last factor is just a $c$-number,
 this can be written as
\begin{equation}
e^{\ahat + \ahat^\dagger} = e^{\frac{1}{2}} \, e^{\ahat^\dagger} e^{\ahat}.
\end{equation}
Note that this is the same argument presented in \cite{Trickle:2019nya} to compute the amplitude for single-phonon production, only restricted here to the greatly simplified context of a single 3-dimensional oscillator. 

Since we are interested in simplifying $e^{i q_x \hat{x}_N} = e^{i \kappa (\ahat + \ahat^\dagger)}$, where $\kappa =i q_x/\sqrt{2 m_N  \omegaO}$, the relevant commutator is $[\kappa a, \kappa a^\dagger] = \kappa^2$. Following the above argument  yields
\begin{equation}
\exp\left[\frac{i q_x}{\sqrt{2 m_N  \omegaO}} \left(\ahat + \ahat^\dagger \right) \right]
= 
\exp\left(-\frac{q_x^2}{4 m_N  \omegaO}\right)
\exp\left(\frac{i q_x}{\sqrt{2 m_N  \omegaO}} \ahat^\dagger \right) 
\exp \left(           \frac{i q_x}{\sqrt{2 m_N  \omegaO}} \ahat   \right)~.
\end{equation}
Consider taking the matrix element of this operator between the states $\langle 0 |$ and $|n_x \rangle$. Acting on $|n_x \rangle$ on the left, we have to get to the state $|0 \rangle$ to have a nonzero matrix element with $\langle 0 |$. The only way to get there is to act $n_x$ times with $\ahat$ and zero times with $\ahat^\dagger$. This means we take the $n_x^{\rm th}$ term from the right-most exponential series, and the $0^{\rm th}$ term from the middle exponential. 
Since these operators satisfy 
\begin{equation}
\ahat^n_x |n_x \rangle = \sqrt{n_x!} | 0 \rangle,
\end{equation}
we can act on our initial and final states to obtain 
\begin{equation}
\left \langle 0 \left | 
\exp\left(-\frac{q_x^2}{4 m_N  \omegaO}\right)
\exp\left(\frac{i q_x}{\sqrt{2 m_N  \omegaO}} \ahat^\dagger \right) 
\exp \left(           \frac{i q_x}{\sqrt{2 m_N  \omegaO}} \ahat   \right)
\right | n_x \right \rangle =  \frac{\sqrt{n_x!}}{n_x!} \left(\frac{i q_x}{\sqrt{2 m_N  \omegaO}}\right)^{n_x} 
\exp\left(-\frac{q_x^2}{4 m_N  \omegaO}\right)~.
\end{equation}
Taking the modulus squared of this expression gives
\begin{equation}
\label{eq:Pnx}
|\langle 0 | e^{i q_x \hat{x}_N} | n_x \rangle |^2 = \frac{1}{n_x !} \left(\frac{q_x}{\sqrt{2 m_N  \omegaO}}\right)^{2n_x}\exp\left(-\frac{q_x^2}{2 m_N  \omegaO}\right),
\end{equation}
and multiplying identical expression for the $n_y$ and $n_z$ contributions recovers~(\ref{eq:PN}).  In fact, from~\Eq{eq:Pnx} we can see that the Cartesian occupation numbers $n_x$, $n_y$, and $n_z$ are also Poisson-distributed, so their sum $n = n_x + n_y + n_z$ will also be Poisson, confirming our more detailed calculation.

  \section{3. Generalizing the Migdal Effect}
Dark matter scattering off bound nuclear targets can also yield electronic energy in the
form of Migdal ionization. However, unlike previous studies of this effect \cite{Ibe:2017yqa}, here the  
nucleus is not a free particle, so here we revisit and generalize this result with a harmonic oscillator  
dispersion relation for the nuclear target. Rather than considering the final state in the boosted frame of the recoiling nucleus, we will decompose the problem into relative coordinates in the lab frame, which more easily generalizes for a bound nucleus.\footnote{We thank Gordon Baym for suggesting this perspective on the problem.}

\subsection{A. Choosing Coordinates}
 We begin by first considering a simple atomic system in which
a single electron and nucleus are  bound by a potential $V_e$,
and the nucleus is held in place by a harmonic oscillator potential. The Hamiltonian
for this system 
can be written
\be
\hat{H} = \frac{\hat{\pp}_N^2}{2 m_N} + \frac{\hat{\pp}_e^2}{2m_e} +\frac{m_N \omegaO^2 }{2} \hat{\rr}_N^2 +V_e(\hat{\rr}_N - \hat{\rr}_e).
\ee
Transforming to relative and center-of-mass coordinates,
\be
\hat{\rr}  = \hat{\rr}_N - \hat{\rr}_e, ~~~
\hat{\RR}  = \frac{  m_N \hat{\rr}_N + m_e \hat{\rr}_e }{m_N + m_e}, 
\ee
the Hamiltonian becomes $\hat{H} = \hat{H}_0 + \Delta \hat{H}$, where  
\be
\hat{H}_0  &=& \frac{\hat{\pp}_R^2}{2(m_N+m_e)} + \frac{\hat{\pp}_r^2}{2\mu} +\frac{m_N \omegaO^2}{2}  \hat{\RR}^2 +V_e(\hat{\rr}),~~
\ee
where $\mu = m_e m_N(m_e + m_N) \approx m_e$ is the electron-nucleus reduced mass, and expanding the original harmonic oscillator term gives
\be
\Delta \hat{H}  &=& -\mu \omegaO^2 \hat{\RR} \cdot \hat{\rr} +  \frac{\mu^2 \omegaO^2}{2m_N}\, \hat{\rr}^2~.
\ee
Written in this way, $\hat{H}_0$ is separable and can be solved by $\Psi(\RR, \rr) = \psi_N(\RR) \psi_e(\rr)$, where $\psi_N$ is a simple harmonic oscillator wavefunction for the nucleus $N$ and $\psi_e$ an electronic wavefunction. The terms in $\Delta \hat{H}$ are suppressed by powers of $m_e/m_N \ll 1$ and can be treated as small perturbations. In particular, the first-order energy shift is parametrically
\be
\label{eq:Eshift}
\Delta E = \langle \Psi | \Delta \hat{H} | \Psi \rangle \sim \frac{m_e^2}{m_N}\omegaO^2 a^2~,
\ee
where we have assumed that the typical spread in position space of the electronic wavefunction is of order the lattice spacing $a$; note that $\Delta E$ independent of the harmonic oscillator level $N$ because $\langle \hat{\RR} \rangle = 0$ in any stationary state, so only the second term in $\Delta \hat{H}$ contributes. For $\omegaO \sim 50 \ {\rm meV}$ and $m_N= 28 \ {\rm GeV}$ as for silicon, we have $\Delta E \sim 25 \ {\rm neV} \ll \omegaO$ and thus we are justified in ignoring the perturbation and treating the nuclear spectrum as purely a harmonic oscillator spectrum.

\subsection{B. Including Dark Matter}

Equipped with this formalism, we can now include the contact interaction from \Eq{VDM} that couples
the DM to the nucleus and 
repeat the argument that culminates in
\Eq{eq:RtotSM} in Sec. 1 of this supplement with the initial/final states 
\be
|\Psi \rangle &=& \sqrt{2m_\chi} \sqrt{2m_N} \,  \psi_0 (\mathbf{r}_N)  \psi_e(\rr_e) \, e^{i \mathbf{p} \cdot\mathbf{r }_\chi}  \\ 
|\Psi^\prime \rangle &=& \sqrt{2m_\chi} \sqrt{2m_N}   \, \psi_n (\mathbf{r}_N) \psi_e^\prime(\rr_e)  \, e^{i \mathbf{p}^\prime \cdot\mathbf{r}_\chi} ,
\ee
where we have merely extended \Eq{psis} to include electron wave functions which have non-relativistic normalization to match the conventions of \cite{Ibe:2017yqa}. Since the DM potential
is only a function of $\rr_N - \rr_\chi$, all the steps leading up to  \Eq{eq:MNf}  are identical 
and the $\psi_e$ states are spectators up until the last step
where, instead of $\langle \vec{n} |   e^{i \qq\cdot \hat{\rr}_N}   | 0 \rangle$,
we get 
\be
\langle \Psi' | e^{i \qq \cdot \hat{\rr}_N} |\Psi \rangle = \langle \vec{n} | e^{i \qq \cdot \hat{\RR}} | 0 \rangle \langle \psi_{e}^\prime | e^{i \qq_e \cdot \hat{\rr}} | \psi_e \rangle,~~~\qq_e \equiv \left(\frac{m_e}{m_N} \right) \qq~.
\ee
Upon squaring this result, the first factor recovers \Eq{eq:PN} and the
second factor  is the same ionization probability
found in Ref. \cite{Ibe:2017yqa}:
\be
Z_f(\qq_e)  \equiv  \langle   \psi_e^\prime | e^{i\qq_e \cdot \hat \rr}  | \psi_e \rangle .
\ee
Although here we have only considered single-electron atoms, the calculation straightforwardly generalizes to atomic systems
with multiple electrons at relative coordinates $\rr_j$:
\be
Z_f(\qq_e)\to \int     \prod_j d^3   \rr_j \psi_e^{\prime *}(\rr_j) \,  \exp \biggl(    i \sum_j \qq_e \cdot  \rr_j  \biggr)  \psi_e(\rr_j).
\ee
Thus, including the electronic matrix elements, the total scattering rate for the Migdal effect now becomes
\be
\label{eq:RtotMig}
R = N_T \frac{\rho_\chi}{m_\chi} \frac{Z^2 \sigmabar_p}{8\pi \mu_{\chi p}^2} \int \frac{d^3 \mathbf  \qq}{q} \sum_n |f(n, \qq)|^2 \, \sum_f |Z_f(\qq_e)|^2  \, \eta(v_{\rm min}^{n,e})~,
\ee
where the sum on $f$ includes all allowed electron final states and
 the additional electronic energy $E_e$ released in this inelastic process shifts
 the  minimum velocity required to undergo a given transition 
\be
v_{\rm min}^{(n,e)} = \frac{E_e + n \omegaO}{q} + \frac{q}{2m_\chi}.
\ee
For systems with spherically symmetric ionization probabilities,
 $Z_f$ depends only on $E_e$ (as in \cite{Ibe:2017yqa} with free atoms), so the electron term factorizes
 from 
the $\frac{1}{4\pi}\int d \Omega_{\qq}$ angular average that yields the Poisson distribution
in \Eq{eq:fNq}. In this case, the differential scattering rate becomes
\be
\frac{dR}{dq dE_e} =  
\! \frac{N_T \rho_\chi}{m_\chi} \frac{Z^2 \sigmabar_p }{2 \mu_{\chi p}^2}
 \sum_n q |f(n, q)|^2 
\sum_f \frac{d|Z_f|^2}{dE_e}
\eta(v_{\rm min}^{n,e}),~~~~~~
\ee
which recovers the expression in \Eq{dRmigdal}. However, in general, it need not
be the case that these terms factorize (in particular, the valence electrons in semiconductors are not necessarily in spherically-symmetric states) and the final distribution will involve
an integration over the combined oscillator and ionization probabilities. 
For a detailed discussion of the Migdal effect in semiconductors see \cite{toappear}.

\end{document}